\documentclass[amssymb,showpacs,aps,prc,floatfix,reprint,superscriptaddress]{revtex4-2}
\usepackage{graphicx}
\usepackage{tabulary}
\usepackage{amssymb}
\usepackage{longtable}
\usepackage{filecontents}
\usepackage{epsf}
\usepackage{epstopdf}
\usepackage{threeparttable}
\usepackage{amsmath}
\usepackage{array}
\usepackage{multirow}
\usepackage{bm}
\usepackage{xcolor}
\usepackage{dcolumn}
\begin{document}
\makeatletter
\newcolumntype{B}[3]{>{\boldmath\DC@{#1}{#2}{#3} }c<{\DC@end} }
\makeatother

\title{Emerging collectivity from the nuclear structure of $^{132}$Xe: Inelastic neutron scattering studies and shell-model calculations}

\author{E.~E.~Peters}
\altaffiliation{email: fe.peters@uky.edu}
\affiliation{Department of Chemistry, University of Kentucky, Lexington, Kentucky 40506-0055, USA}
\author{A.~E.~Stuchbery}
\affiliation{Department of Nuclear Physics, Research School of Physics and Engineering, The Australian National University, Canberra ACT 2601, Australia}
\author{A.~Chakraborty}
\altaffiliation{Present Address: Department~of~Physics,~Siksha~Bhavana,~Visva-Bharati, Santiniketan~731~235,~West~Bengal,~India}
\affiliation{Department of Chemistry, University of Kentucky, Lexington, Kentucky 40506-0055, USA}
\affiliation{Department of Physics \& Astronomy, University of Kentucky, Lexington, Kentucky 40506-0055, USA}
\author{B.~P.~Crider}
\altaffiliation{Present Address:Department of Physics and Astronomy, Mississippi State University, Mississippi State, Mississippi 39762, USA }
\affiliation{Department of Physics \& Astronomy, University of Kentucky, Lexington, Kentucky 40506-0055, USA}
\author{S.~F.~Ashley}
\affiliation{Department of Chemistry, University of Kentucky, Lexington, Kentucky 40506-0055, USA}
\affiliation{Department of Physics \& Astronomy, University of Kentucky, Lexington, Kentucky 40506-0055, USA}
\author{A.~Kumar}
\altaffiliation{Present Address: Department~of~Physics,~Banaras~Hindu~University,~Varanasi~221005~India}
\affiliation{Department of Physics \& Astronomy, University of Kentucky, Lexington, Kentucky 40506-0055, USA}
\affiliation{Department of Chemistry, University of Kentucky, Lexington, Kentucky 40506-0055, USA}
\author{M.~T.~McEllistrem}
\affiliation{Department of Physics \& Astronomy, University of Kentucky, Lexington, Kentucky 40506-0055, USA}
\author{F.~M.~Prados-Est\'{e}vez}
\affiliation{Department of Chemistry, University of Kentucky, Lexington, Kentucky 40506-0055, USA}
\affiliation{Department of Physics \& Astronomy, University of Kentucky, Lexington, Kentucky 40506-0055, USA}
\author{S.~W.~Yates}
\altaffiliation{email: yates@uky.edu}
\affiliation{Department of Chemistry, University of Kentucky, Lexington, Kentucky 40506-0055, USA}
\affiliation{Department of Physics \& Astronomy, University of Kentucky, Lexington, Kentucky 40506-0055, USA}

\date{\today}

\begin{abstract}

Inelastic neutron scattering was used to study the low-lying nuclear structure of $^{132}$Xe.  A comprehensive level scheme is presented, as well as new level lifetimes, multipole mixing ratios, branching ratios, and transition probabilities.  Comparisons of these data as well as previously measured $E2$ strengths and $g$ factors are made with new shell-model calculations for $^{132,134,136}$Xe to explore the emergence of collectivity in the Xe isotopes with $N<82$ near the closed shell.

\end{abstract}

\maketitle

\section{Introduction}

How do ``simple'' collective motions emerge from the complexity of the underlying nucleon-nucleon interactions?  To answer this question, studies of isotopic chains that progress from a semi-magic nuclide towards isotopes with vibrational and then rotational-like structures form an important landscape. The isotopes nearest closed shells are particularly important from a microscopic perspective because the first signals of the emergence of collectivity can be studied through large-basis shell-model calculations.

The nine ``stable'' isotopes of Xe ($A$ = 124, 126, 128, 129, 130, 131, 132, 134, 136) span a transitional region of nuclear structure that has yet to be fully characterized.  While the light-mass isotopes appear to be gamma-soft rotors \cite{Rad15} and $^{136}$Xe at the closed $N=82$ shell exhibits seniority structure \cite{Xe136.PhysRevC.98.034302}, the nature of those in between is not well understood.  Certainly, collectivity is emerging as the number of neutron holes increases away from $^{136}$Xe. Moreover, as will be discussed, the seniority structure of the proton configuration in $^{136}$Xe makes the $E2$ transition strengths of the xenon isotopes particularly sensitive to the emergence of collectivity. 

In our previous work on $^{130,132}$Xe \cite{Pet16}, we sought a comparison with the E(5) critical-point symmetry, for which $^{130}$Xe had been proposed a candidate \cite{Coquard128}.  However, neither nucleus emerged as a clear-cut representation of that symmetry.  That publication \cite{Pet16} included only a truncated level scheme (up to 2.2 MeV) for both nuclei relevant to the E(5) depiction, but we have now fully analyzed the more extensive data set (up to 3.3 MeV) for $^{132}$Xe. Previous measurements for $^{132}$Xe have yielded limited data or are several decades old.  The most extensive data sets are those from $\beta$-decay measurements, but the most recent of these was published in 1982 \cite{Soo83}.  The Nuclear Data Sheets (NDS) compilation \cite{NDS132} for the $\beta^-$ decay of $^{132}$I relies most heavily on the work in Refs.~\cite{Net78, Sin73, Gir80}, which was carried out using Ge(Li) and/or NaI detectors.  In addition, transfer reactions \cite{Ker71, Ott96} were performed more than twenty years ago, as well as neutron capture \cite{Gro71, Gel71, Ham88}.  The most current data come from Coulomb excitation \cite{Coq10} and photon scattering measurements \cite{von06}, which populate limited selections of states.  A study of $^{132}$Xe affords an opportunity for inelastic neutron scattering (INS) to develop a more comprehensive picture of the level scheme and to provide level lifetimes from the Doppler-shift attenuation method (DSAM) allowing the determination of reduced transition probabilities.

Along with the new experimental data, we seek insights into the emergence of collectivity at the microscopic level in the xenon isotopes near $^{136}$Xe through large-basis shell-model calculations. A number of shell-model calculations have been performed for the Xe isotopes near $N=82$ \cite{Xepaper.PhysRevC.65.024316,BRN.PhysRevC.71.044317,Teruya.PhysRevC.92.034320,AES-Te136RIV.PhysRevC.96.014321,Vogt.PhysRevC.93.054325,Vogt.PhysRevC.96.024321,Xe136.PhysRevC.98.034302}. Recent work includes an extensive study of nuclei around mass 130 by Teruya \textit{et al.} \cite{Teruya.PhysRevC.92.034320}, calculations on $^{132}$Xe and $^{134}$Xe to high spin by Vogt \textit{et al.} \cite{Vogt.PhysRevC.93.054325,Vogt.PhysRevC.96.024321}, and calculations for the low-seniority states of $^{136}$Xe by Van Isacker \cite{Xe136.PhysRevC.98.034302}. These calculations included all of the orbitals in the $50 \le N,Z \leq 82$ major shell for both protons and neutrons, namely $0g_{7/2}$, $1d_{5/2}$, $1d_{3/2}$, $2s_{1/2}$, and $0h_{11/2}$, but employed different interactions. Van Isacker \cite{Xe136.PhysRevC.98.034302} also performed calculations in a reduced model space of the proton $0g_{7/2}$ and $1d_{5/2}$ orbitals to help identify the seniority structure of the low-lying levels in $^{136}$Xe.

The objective of the present shell-model calculations is to track the emergence of collectivity from a microscopic perspective as the number of neutron holes increases from $^{136}$Xe to $^{132}$Xe by examining patterns in the level structures, increasing $E2$ transition strengths, the magnitudes and ratios of excited-state $g$~factors, and the increasing complexity of the wavefunctions. The xenon isotopes are well suited for such an investigation because the pronounced seniority patterns of the $E2$ transitions in $^{136}$Xe must be ``washed out'' as collectivity develops.


\section{Experiments and Results}

The experiments from which the majority of the current data were extracted, using inelastic neutron scattering from a solid, highly enriched $^{132}$XeF$_2$ sample, were described in Ref. \cite{Pet16}.  An additional angular distribution measurement at an incident neutron energy ($E_n$) of 3.4 MeV was performed and these data are included in the present work.  The prior publication \cite{Pet16} only included a partial level scheme, but we now offer the full version as obtained from our $(n,n^{\prime}\gamma)$ measurements.  A summary of the data for levels in $^{132}$Xe is given in Table \ref{tb:nng}; comments on levels to which these measurements have uniquely contributed are provided.  Angular distributions of $\gamma$ rays to the ground state with a positive value of $a_2$ are described as ``quadrupole", while those with a negative $a_2$ are described as ``dipole".

\setlength{\LTcapwidth}{\textwidth}
\begin{longtable*}{@{\extracolsep{\fill}}ddcccdcccc>{\raggedright}p{4.5in}|p{0.85in}|}

\caption{Data extracted from the present $(n,n^{\prime}\gamma)$ experiments for $^{132}$Xe.  When two mixing ratios are possible, the solution with the lowest $\chi ^2$ value is listed first.  The final column is the reduced transition probability for either $M1$ or $E1$ multipolarity, as appropriate.}\label{tb:nng}\\

\hline
\hline
\noalign{\smallskip}
\multicolumn{1}{c}{$E_{level}$} & \multicolumn{1}{c}{$E_{\gamma}$} & $J^\pi_i$ & $J^\pi_f$ & B.R. & \multicolumn{1}{r}{$\bar{F}(\tau)$} &  $\tau$  &  $\delta$         & B(E2) & B(M1)/B(E1) \\
\multicolumn{1}{c}{(keV)}       & \multicolumn{1}{c}{(keV)}        &           &           &      &           & (fs)       & or multipolarity  & (W.u.)& ($\mu_N^2$)/(mW.u.) \\
\noalign{\smallskip}
\endfirsthead


\caption{(Continued.)}\\
\hline
\hline
\noalign{\smallskip}
\multicolumn{1}{c}{$E_{level}$} & \multicolumn{1}{c}{$E_{\gamma}$} & $J^\pi_i$ & $J^\pi_f$ & B.R. & \multicolumn{1}{r}{$\bar{F}(\tau)$} &  $\tau$  &  $\delta$          & B(E2) & B(M1)/B(E1) \\
\multicolumn{1}{c}{(keV)}       & \multicolumn{1}{c}{(keV)}        &           &           &      &           & (fs)       &  or multipolarity  & (W.u.)& ($\mu_N^2$)/(mW.u.) \\
\noalign{\smallskip}
\hline
\noalign{\smallskip}
\endhead

\noalign{\smallskip}
\hline
\endfoot
\noalign{\smallskip}
\hline
\endlastfoot

\hline
\noalign{\smallskip}

667.716	(	2	)	&	667.714	(	2	)	&	 $2^+_1$ 	&	 $0^+_1$  	&	1	&	   	&	   	&	   	&	   	&	   \\
1297.946	(	4	)	&	630.227	(	4	)	&	 $2^+_2$ 	&	 $2^+_1$  	&	  $^b$ 	&	   	&	   	&	   	&	   	&	   \\
				&	 1297.939^a 				&	 	&	 $0^+_1$ 	&	   	&	   	&	   	&	   	&	   	&	   \\
1440.368	(	5	)	&	772.645	(	5	)	&	 $4^+_1$ 	&	 $2^+_1$  	&	1	&	   	&	   	&	   	&	   	&	   \\
1803.814	(	6	)	&	363.443	(	24	)	&	 $3^+_1$ 	&	 $4^+_1$  	&	 0.048(3)  	&	   	&	   	&	   	&	   	&	   \\
				&	505.869	(	6	)	&	         	&	 $2^+_2$  	&	 0.574(14) 	&	   	&	   	&	   	&	   	&	   \\
				&	1136.064	(	11	)	&	         	&	 $2^+_1$  	&	 0.378(13) 	&	   	&	   	&	   	&	   	&	   \\
1948.207	(	13	)	&	1280.477	(	13	)	&	 $0^+_2$ 	&	 $2^+_1$  	&	1	&	 0.044(32) 	&	 $1500^{+3900}_{-700}$ 	&	 E2                    	&	 $4.0^{+31}_{-29}$   	&	   \\
1962.982	(	9	)	&	522.605	(	7	)	&	 $4^+_2$ 	&	 $4^+_1$  	&	 0.879(6)  	&	 0.047(24) 	&	 $1500^{+1500}_{-500}$ 	&	 $-0.214^{+23}_{-26}$  	&	 $14^{+12}_{-8}$     	&	  $0.22^{+12}_{-11}$  \\
				&	1295.62	(	25	)	&	         	&	 $2^+_1$  	&	 0.121(12)  	&	           	&	                       	&	 E2                    	&	 $0.45^{+29}_{-25}$  	&	   \\
1985.660	(	7	)	&	1317.923	(	8	)	&	 $2^+_3$ 	&	 $2^+_1$  	&	 0.899(7)  	&	 0.531(18) 	&	 63(4)                 	&	 $-0.201^{+26}_{-23}$  	&	 $2.85^{+92}_{-82}$  	&	 $0.341^{+29}_{-26}$ \\
				&	1985.660	(	27	)	&	         	&	 $0^+_1$  	&	 0.101(7)  	&	           	&	                       	&	 E2                    	&	 $1.06^{+15}_{-13}$  	&	   \\
2040.411	(	10	)	&	600.035	(	8	)	&	 $5^-_1$ 	&	 $4^+_1$  	&	1	&	           	&	                       	&	 E1                    	&	                     	&	   \\
2110.240	(	12	)	&	 669.862^a    				&	 $4^+_3$ 	&	 $4^+_1$  	&	 $^b$  	&	   	&	   	&	   	&	   	&	   \\
				&	812.283	(	11	)	&	         	&	 $2^+_2$  	&	   	&	   	&	   	&	 E2  	&	   	&	   \\
				&	 1442.508^a   				&	         	&	 $2^+_1$  	&	   	&	   	&	   	&	 E2  	&	   	&	   \\
2167.369	(	23	)	&	726.991	(	22	)	&	 $6^+_1$ 	&	 $4^+_1$  	&	1	&	 0.091(80) 	&	 $700^{+5300}_{-400}$  	&	 E2                    	&	 $140^{+150}_{-130}$ 	&	   \\
2169.258	(	14	)	&	1501.525	(	14	)	&	 $0^+_3$ 	&	 $2^+_1$  	&	1	&	 0.241(27) 	&	 $229^{+37}_{-30}$     	&	 E2                    	&	 $11.7^{+18}_{-16}$  	&	   \\
2187.424	(	8	)	&	889.464	(	10	)	&	 $2^+_4$ 	&	 $2^+_2$  	&	 0.367(14) 	&	 0.276(16) 	&	 $191^{+15}_{-14}$     	&	 $-0.064^{+48}_{-46}$  	&	 $0.29^{+66}_{-27}$  	&	 $0.155^{+19}_{-18}$ \\
				&	1519.691	(	12	)	&	         	&	 $2^+_1$  	&	 0.549(13) 	&	           	&	                       	&	 $1.50^{+19}_{-17}$    	&	 $5.01^{+92}_{-82}$  	&	 $0.0144^{+42}_{-34}$  \\
				&					&	         	&	          	&	           	&	           	&	                       	&	 $0.197^{+60}_{-55}$   	&	 $0.27^{+23}_{-14}$  	&	 $0.0448^{+56}_{-53}$  \\
				&	2187.53	(	14	)	&	         	&	 $0^+_1$  	&	 0.084(6)  	&	           	&	                       	&	 E2                    	&	 $0.179^{+28}_{-25}$ 	&	   \\
2272.423	(	20	)	&	974.574	(	90	)	&	 $0^+_4$ 	&	 $2^+_2$  	&	 0.243(15) 	&	 0.113(36) 	&	 $560^{+280}_{-150}$   	&	 E2                    	&	 $10.2^{+46}_{-38}$  	&	   \\
				&	1604.682	(	20	)	&	         	&	 $2^+_1$  	&	 0.757(15) 	&	           	&	                       	&	 E2                    	&	 $2.6^{+10}_{-9}$    	&	   \\
2288.221	(	12	)	&	1620.485	(	12	)	&	$(3^+$)  	&	 $2^+_1$  	&	1	&	 0.147(21) 	&	 $413^{+78}_{-58}$     	&	   	&	   	&	   \\
2303.591	(	23	)	&	863.210	(	22	)	&	 $6^+_2$ 	&	 $4^+_1$  	&	1	&	           	&	                       	&	 E2  	&	   	&	   \\
2306.658	(	20	)	&	343.659	(	19	)	&	$(4^+)$  	&	 $4^+_2$  	&	 0.449(13) 	&	 0.40(12)  	&	 $110^{+73}_{-40}$     	&	   	&	   	&	   \\
				&	866.325	(	69	)	&	         	&	 $4^+_1$  	&	 0.165(13) 	&	           	&	                       	&	   	&	   	&	   \\
				&	1008.87	(	12	)	&	         	&	 $2^+_2$  	&	 0.319(12) 	&	           	&	                       	&	 (E2)  	&	   	&	   \\
				&	1639.04	(	12	)	&	         	&	 $2^+_1$  	&	 0.067(7)  	&	           	&	                       	&	 (E2)  	&	   	&	   \\
2350.734	(	20	)	&	546.904	(	23	)	&	 $5^+_1$ 	&	 $3^+_1$  	&	 0.538(27) 	&	           	&	                       	&	 E2  	&	   	&	   \\
				&	910.370	(	37	)	&	         	&	 $4^+_1$  	&	 0.462(27) 	&	           	&	                       	&	 $-0.59^{+19}_{-51}$ 	&	   	&	   \\
2353.160	(	26	)	&	312.743	(	24	)	&	 $(4,6)^-$  	&	 $5^-_1$  	&	1	&	   	&	   	&	   	&	   	&	   \\
2387.924	(	14	)	&	1720.184	(	14	)	&	 $2^+_5$ 	&	 $2^+_1$  	&	 0.918(7)  	&	 0.160(23) 	&	 $371^{+72}_{-55}$     	&	 $-4.2^{+9}_{-11}$     	&	 $3.18^{+66}_{-63}$  	&	 $0.0015^{+13}_{-7}$ \\
				&					&	         	&	          	&	           	&	           	&	                       	&	 $-0.74^{+8}_{-12}$    	&	 $1.19^{+49}_{-35}$  	&	 $0.0179^{+50}_{-46}$  \\
				&	2388.003	(	81	)	&	         	&	 $0^+_1$  	&	 0.082(7)  	&	           	&	                       	&	 E2                    	&	 $0.058^{+16}_{-14}$ 	&	   \\
2394.973	(	14	)	&	954.590	(	13	)	&	$(4^+)$  	&	 $4^+_1$  	&	1	&	 0.194(35) 	&	 $295^{+79}_{-55}$     	&	                       	&	                     	&	   \\
2424.823	(	13	)	&	621.004	(	12	)	&	$(3^+)$  	&	 $3^+_1$  	&	 0.668(12) 	&	 0.108(39) 	&	 $580^{+360}_{-170}$   	&	   	&	   	&	   \\
				&	984.360	(	56	)	&	         	&	 $4^+_1$  	&	 0.244(12) 	&	   	&	   	&	   	&	   	&	   \\
				&	1756.80	(	14	)	&	         	&	 $2^+_1$  	&	 0.088(7)  	&	   	&	   	&	   	&	   	&	   \\
2442.536	(	33	)	&	402.118	(	32	)	&	$(4,6)^-$	&	 $5^-_1$	&	1	&	   	&	   	&	   	&	   	&	   \\
2453.960	(	15	)	&	1155.984	(	22	)	&	 $2^+_6$ 	&	 $2^+_2$  	&	 0.399(10) 	&	 0.121(27) 	&	 $510^{+170}_{-100}$   	&	 $-0.157^{+78}_{-83}$  	&	 $0.19^{+36}_{-15}$  	&	 $0.0283^{+89}_{-82}$  \\
				&					&	         	&	          	&	           	&	           	&	                       	&	 $4.1^{+17}_{-10}$     	&	 $7.4^{+24}_{-22}$   	&	 $0.0017^{+19}_{-10}$  \\
				&	1786.229	(	20	)	&	         	&	 $2^+_1$  	&	 0.454(10) 	&	           	&	                       	&	 $1.49^{+26}_{-24}$    	&	 $0.70^{+28}_{-24}$  	&	 $0.0028^{+17}_{-12}$  \\
				&					&	         	&	          	&	           	&	           	&	                       	&	 $0.204^{+87}_{-78}$   	&	 $0.040^{+61}_{-29}$ 	&	 $0.0086^{+27}_{-25}$  \\
				&	2454.11	(	16	)	&	         	&	 $0^+_1$  	&	 0.147(8)  	&	           	&	                       	&	 E2                    	&	 $0.067^{+22}_{-19}$ 	&	   \\
2466.530	(	30	)	&	1798.791	(	30	)	&	 $0^+_5$ 	&	 $2^+_1$  	&	1	&	   	&	   	&	 E2  	&	   	&	   \\
2469.173	(	14	)	&	483.490	(	17	)	&	 $3^-_1$ 	&	 $2^+_3$  	&	 0.463(15) 	&	 0.120(71) 	&	 $510^{+820}_{-210}$   	&	 E1                    	&	                     	&	 $3.0^{+23}_{-19}$ \\
				&	1028.823	(	26	)	&	         	&	 $4^+_1$  	&	 0.276(14) 	&	           	&	                       	&	 E1                    	&	                     	&	 $0.19^{+15}_{-12}$  \\
				&	1801.428	(	43	)	&	         	&	 $2^+_1$  	&	 0.261(10) 	&	           	&	                       	&	 E1                    	&	                     	&	 $0.033^{+26}_{-21}$ \\
2512.040	(	15	)	&	471.620	(	12	)	&	 $4^-$   	&	 $5^-_1$  	&	1	&	   	&	   	&	   	&	   	&	   \\
2526.147	(	26	)	&	1228.093	(	53	)	&	         	&	 $2^+_2$  	&	 0.163(34) 	&	 0.132(42) 	&	 $460^{+250}_{-130}$ 	&	   	&	   	&	   \\
				&	1858.434	(	29	)	&	         	&	 $2^+_1$  	&	 0.837(34) 	&	   	&	   	&	   	&	   	&	   \\
2555.674	(	16	)	&	569.990	(	17	)	&	 $3^-_2$ 	&	 $2^+_3$  	&	 0.530(15) 	&	   	&	   	&	 E1  	&	   	&	   \\
				&	1887.972	(	30	)	&	         	&	 $2^+_1$  	&	 0.470(15) 	&	   	&	   	&	 E1  	&	   	&	   \\
2563.204	(	25	)	&	1895.446	(	28	)	&	1	&	 $2^+_1$  	&	 0.625(12) 	&	 0.527(33) 	&	 $72^{+10}_{-9}$ 	&	   	&	   	&	   \\
				&	2563.239	(	54	)	&	         	&	 $0^+_1$  	&	 0.375(12) 	&	   	&	   	&	   	&	   	&	   \\
2584.098	(	44	)	&	780.286	(	52	)	&	         	&	 $3^+_1$  	&	 0.435(58) 	&	   	&	   	&	   	&	   	&	   \\
				&	1143.674	(	82	)	&	         	&	 $4^+_1$  	&	 0.565(58) 	&	   	&	   	&	   	&	   	&	   \\
2588.754	(	30	)	&	1290.781	(	44	)	&	$2^+,4^+$	&	 $2^+_2$  	&	 0.472(30) 	&	 0.152(91) 	&	 $390^{+690}_{-170}$ 	&	   	&	   	&	   \\
				&	1921.019	(	41	)	&	         	&	 $2^+_1$  	&	 0.528(30) 	&	   	&	   	&	   	&	   	&	   \\
2593.067	(	54	)	&	1152.711	(	59	)	&	         	&	 $4^+_1$  	&	 0.638(28) 	&	   	&	   	&	   	&	   	&	   \\
				&	1925.15	(	14	)	&	         	&	 $2^+_1$  	&	 0.362(28) 	&	   	&	   	&	   	&	   	&	   \\
2594.00	(	11	)	&	1926.23	(	13	)	&	$1,2^+$  	&	 $2^+_1$  	&	 0.373(32) 	&	 0.125(69) 	&	 $480^{+680}_{-190}$ 	&	   	&	   	&	   \\
				&	2594.02	(	18	)	&	         	&	 $0^+_1$  	&	 0.627(32) 	&	   	&	   	&	   	&	   	&	   \\
2613.589	(	50	)	&	650.541	(	60	)	&	 $5^+_2$ 	&	 $4^+_2$  	&	 0.552(35) 	&	           	&	                       	&	 $-5^{+2}_{-11}$       	&	   	&	   \\
				&					&	         	&	          	&	           	&	           	&	                       	&	 $-0.14^{+16}_{-18}$   	&	   	&	   \\
				&	809.880	(	90	)	&	         	&	 $3^+_1$  	&	 0.448(35) 	&	           	&	                       	&	 E2  	&	   	&	   \\
2622.066	(	31	)	&	1324.082	(	39	)	&	$(2^+)$  	&	 $2^+_2$  	&	 0.554(21) 	&	 0.395(68) 	&	 $120^{+40}_{-28}$     	&	   	&	   	&	   \\
				&	1954.384	(	57	)	&	         	&	 $2^+_1$  	&	 0.312(17) 	&	   	&	   	&	   	&	   	&	   \\
				&	2621.93	(	12	)	&	         	&	 $0^+_1$  	&	 0.134(13) 	&	   	&	   	&	 (E2)  	&	   	&	   \\
2670.018	(	23	)	&	1372.050	(	23	)	&	 $3^+$   	&	 $2^+_2$  	&	 $^b$      	&	 0.362(53) 	&	 $137^{+35}_{-26}$     	&	    	&	   	&	   \\
				&	 2002.275^a   				&	         	&	 $2^+_1$  	&	           	&	           	&	                       	&	    	&	   	&	   \\
2693.970	(	29	)	&	2026.226	(	29	)	&	 $3^-$   	&	 $2^+_1$  	&	1	&	 0.082(40) 	&	 $830^{+860}_{-290}$   	&	 E1                    	&	                     	&	 $0.055^{+30}_{-28}$ \\
2713.942	(	30	)	&	1415.973	(	42	)	&	$1^+,2^+$	&	 $2^+_2$  	&	 0.268(11) 	&	 0.788(43) 	&	 $22^{+6}_{-5}$  	&	   	&	   	&	   \\
				&	2046.38	(	14	)	&	         	&	 $2^+_1$  	&	 0.188(10) 	&	   	&	   	&	   	&	   	&	   \\
				&	2713.894	(	44	)	&	         	&	 $0^+_1$  	&	 0.544(13) 	&	   	&	   	&	   	&	   	&	   \\
2721.498	(	36	)	&	2053.754	(	36	)	&	 $2^+$   	&	 $2^+_1$  	&	  $^b$     	&	 0.161(46) 	&	 $390^{+190}_{-100}$   	&	 $-0.03^{+15}_{-13}$ 	&	   	&	   \\
				&					&	         	&	          	&	           	&	           	&	                       	&	 $2.6^{+16}_{-8}$  	&	   	&	   \\
				&	 2721.468^a   				&	         	&	 $0^+_1$  	&	           	&	           	&	                       	&	 E2  	&	   	&	   \\
2754.558	(	59	)	&	 791.561^a    				&	 $4^+$   	&	 $4^+_2$  	&	 $^b$  	&	   	&	   	&	   	&	   	&	   \\
				&	 1456.588^a   				&	         	&	 $2^+_2$  	&	   	&	   	&	   	&	 E2  	&	   	&	   \\
				&	2086.813	(	59	)	&	         	&	 $2^+_1$  	&	   	&	   	&	   	&	 E2  	&	   	&	   \\
2758.094	(	51	)	&	2090.363	(	84	)	&	 $2^+$   	&	 $2^+_1$  	&	 0.506(20) 	&	   	&	   	&	   	&	   	&	   \\
				&	2758.055	(	65	)	&	         	&	 $0^+_1$  	&	 0.494(20) 	&	   	&	   	&	 E2  	&	   	&	   \\
2781.327	(	43	)	&	1483.62	(	17	)	&	         	&	 $2^+_2$  	&	 0.482(19) 	&	 0.488(84) 	&	 $82^{+32}_{-23}$  	&	   	&	   	&	   \\
				&	2113.563	(	44	)	&	         	&	 $2^+_1$  	&	 0.518(19) 	&	   	&	   	&	   	&	   	&	   \\
2818.527	(	37	)	&	2150.780	(	37	)	&	$(3^-)$  	&	 $2^+_1$  	&	1	&	 0.256(47) 	&	 $220^{+65}_{-45}$ 	&	 (E1)  	&	   	&	   \\
2821.12	(	33	)	&	2821.09	(	33	)	&	 $1,2^+$ 	&	 $0^+_1$  	&	1	&	   	&	   	&	   	&	   	&	   \\
2839.372	(	59	)	&	1398.980	(	59	)	&	         	&	 $4^+_1$  	&	1	&	 0.31(12)  	&	 $170^{+140}_{-70}$  	&	   	&	   	&	   \\
2840.191	(	35	)	&	877.189	(	35	)	&	         	&	 $4^+_2$  	&	 0.796(25) 	&	 0.27(10)  	&	 $210^{+170}_{-80}$  	&	   	&	   	&	   \\
				&	2172.51	(	18	)	&	         	&	 $2^+_1$  	&	 0.204(25) 	&	   	&	   	&	   	&	   	&	   \\
2872.729	(	41	)	&	832.301	(	40	)	&	$(4,6)^-$	&	 $5^-_1$  	&	1	&	   	&	   	&	   	&	   	&	   \\
2890.748	(	50	)	&	927.747	(	49	)	&	         	&	 $4^+_2$  	&	1	&	 0.17(14)  	&	 $400^{+1900}_{-200}$  	&	   	&	   	&	   \\
2896.633	(	44	)	&	1598.662	(	45	)	&	$(3^+)$  	&	 $2^+_2$  	&	 0.738(98) 	&	 0.222(97) 	&	 $260^{+250}_{-100}$ 	&	   	&	   	&	   \\
				&	2228.87	(	16	)	&	         	&	 $2^+_1$  	&	 0.262(98) 	&	   	&	   	&	   	&	   	&	   \\
2903.026	(	46	)	&	1099.059	(	83	)	&	$2^+,4^+$	&	 $3^+_1$  	&	 0.444(27) 	&	 0.324(65) 	&	 $158^{+57}_{-38}$ 	&	   	&	   	&	   \\
				&	2235.335	(	55	)	&	         	&	 $2^+_1$  	&	 0.556(27) 	&	   	&	   	&	   	&	   	&	   \\
2916.95	(	12	)	&	1476.56	(	12	)	&	         	&	 $4^+_1$  	&	1	&	   	&	   	&	   	&	   	&	   \\
2922.293	(	62	)	&	2922.258	(	62	)	&	1	&	 $0^+_1$  	&	1	&	 0.274(56) 	&	 $199^{+69}_{-45}$ 	&	   	&	   	&	   \\
2928.902	(	65	)	&	2261.03	(	16	)	&	 $2^+$   	&	 $2^+_1$  	&	 0.238(26) 	&	 0.385(61) 	&	 $122^{+36}_{-26}$     	&	                       	&	 $0.68^{+28~c}_{-22}$  	&	 $0.0096^{+39~c}_{-30}$  \\
				&	2928.892	(	72	)	&	         	&	 $0^+_1$  	&	 0.762(26) 	&	           	&	                       	&	 E2                    	&	 $0.59^{+19}_{-15}$  	&	   \\
2959.99	(	19	)	&	2959.95	(	19	)	&	$1,2^+$  	&	 $0^+_1$  	&	1	&	   	&	   	&	   	&	   	&	   \\
2968.995	(	47	)	&	1671.030	(	50	)	&	$1,2^+$  	&	 $2^+_2$  	&	 0.602(25) 	&	 0.294(85) 	&	 $180^{+100}_{-60}$  	&	   	&	   	&	   \\
				&	2301.22	(	16	)	&	         	&	 $2^+_1$  	&	 0.252(23) 	&	   	&	   	&	   	&	   	&	   \\
				&	2968.81	(	22	)	&	         	&	 $0^+_1$  	&	 0.146(20) 	&	   	&	   	&	   	&	   	&	   \\
3050.826	(	76	)	&	2383.049	(	84	)	&	 $2^+$   	&	 $2^+_1$  	&	 0.749(29) 	&	 0.232(81) 	&	 $250^{+170}_{-80}$    	&	                       	&	 $0.81^{+45~c}_{-35}$  	&	 $0.0128^{+71~c}_{-55}$  \\
				&	3050.91	(	18	)	&	         	&	 $0^+_1$  	&	 0.251(29) 	&	           	&	                       	&	 E2                    	&	 $0.079^{+53}_{-38}$ 	&	   \\
3058.10	(	17	)	&	1760.12	(	17	)	&	$(3^+)$  	&	 $2^+_2$  	&	 $^b$  	&	   	&	   	&	   	&	   	&	   \\
				&	 2390.35^a    				&	         	&	 $2^+_1$  	&	   	&	   	&	   	&	   	&	   	&	   \\
3076.586	(	72	)	&	1272.747	(	72	)	&	         	&	 $3^+_1$  	&	1	&	   	&	   	&	   	&	   	&	   \\
3091.640	(	75	)	&	1793.67	(	13	)	&	1	&	 $2^+_2$  	&	 0.239(30) 	&	 0.619(99) 	&	 $48^{+23}_{-17}$  	&	   	&	   	&	   \\
				&	2423.86	(	18	)	&	         	&	 $2^+_1$  	&	 0.273(52) 	&	   	&	   	&	   	&	   	&	   \\
				&	3091.61	(	11	)	&	         	&	 $0^+_1$  	&	 0.487(44) 	&	   	&	   	&	   	&	   	&	   \\
3113.224	(	65	)	&	2445.457	(	70	)	&	 $2^+$   	&	 $2^+_1$  	&	 0.760(24) 	&	 0.608(68) 	&	 $50^{+15}_{-13}$      	&	                       	&	 $3.6^{+14~c}_{-10}$   	&	 $0.059^{+23~c}_{-15}$ \\
				&	3113.27	(	18	)	&	         	&	 $0^+_1$  	&	 0.240(24) 	&	           	&	                       	&	 E2                    	&	 $0.34^{+16}_{-10}$  	&	   \\
3145.50	(	43	)	&	2477.75	(	43	)	&	         	&	 $2^+_1$  	&	1	&	   	&	   	&	   	&	   	&	   \\
3274.40	(	15	)	&	2606.64	(	15	)	&	         	&	 $2^+_1$  	&	1	&	   	&	   	&	   	&	   	&	   \\

\end{longtable*}
 \onecolumngrid
 \noindent $^a$ $E_{\gamma}$ was determined from non-recoil-corrected level energy differences due to contamination from other origins.\\
 \noindent $^b$ Branching ratios could not be determined due to contamination from other origins.\\
 \noindent $^c$ This value was calculated assuming pure $E2$ or $M1$ multipolarity.\\
 \twocolumngrid

\subsection{Newly observed levels}

1948.2 keV $0^+_2$ state.  As described in Ref. \cite{Pet16}, this state was identified for the first time in our INS measurements.

2272.4 keV $0^+_4$ state.  The angular distributions of both the 974.6 and 1604.7 keV $\gamma$ rays are isotropic, and the level cross section from the excitation function matches well with a spin-parity of $0^+$.

2288.2 keV $(3^+)$ level.  The threshold of the 1620.5 keV $\gamma$ ray is 2.3 MeV, and the angular distribution indicates either spin 2 or 3 with a measurable mixing ratio.  No ground-state transition is observed, therefore, the spin-parity is tentatively assigned as $(3^+)$.

2306.6 keV $(4^+)$ level.  Four $\gamma$ rays are placed from this level as transitions to the $2^+_1$, $2^+_2$, $4^+_1$, and $4^+_2$ states.  Based on the angular distributions of the 343.7 and 1008.9 keV $\gamma$ rays, the spin is tentatively assigned as $(4^+)$.

2387.9 keV $2^+_5$ state.  The 2388.0 keV $\gamma$ ray has a 2.5 MeV threshold and a quadrupole angular distribution, which leads to the conclusion that it is a $2^+ \rightarrow 0^+_1$ transition.

2442.5 keV $(4,6)^-$ level.  The 402.1 keV $\gamma$ ray has a threshold of 2.6 MeV and is placed as feeding the $5^-_1$ state.  The angular distribution indicates either spin 4 or 6 with a measurable mixing ratio, which leads to the conclusion of negative parity.

2453.9 keV $2^+_6$ state.  A 2454 keV $\gamma$ ray with a 2.5 MeV threshold is observed to have a quadrupole angular distribution, establishing it as a $2^+ \rightarrow 0^+_1$ transition.  Transitions to the $2^+_1$ and $2^+_2$ states are also placed.

2466.5 keV $0^+_5$ state.  At 2.5 MeV, a 1798.8 keV $\gamma$ ray is observed and has an isotropic angular distribution.  The excitation function when compared with \textsc{cindy} \cite{cindy} calculations also agrees with a spin-parity of $0^+$.

2526.1 keV level.  This level is based on the observation of a 1858.4 keV $\gamma$ ray with a 2.6 MeV threshold.  The angular distribution does not provide enough information to assign a spin or parity.  A much weaker 1228.1 keV $\gamma$ ray is also placed from this level as a transition to the $2^+_2$ state.

2563.2 keV 1 level.  A 2563.3 keV $\gamma$ ray was observed with a 2.7 MeV threshold, indicating a ground-state transition.  The angular distribution is dipole in shape, implying the spin is 1, but the parity could not be deduced.  A $\gamma$ ray to the $2^+_1$ state is also observed, but with a nondescript angular distribution.

2593.0 keV level.  This level and the following one are based on a doublet of $\gamma$ rays at 1925.2 and 1926.2 keV.  A $\gamma$ ray to the $4^+_1$ state is also observed, but is too weak to determine the spin and parity.

2594.0 keV $1,2^+$ level.  In addition to the 1926.2 keV $\gamma$ ray, a $\gamma$ ray to the ground state is also observed, but its angular distribution does not allow a distinction between spin 1 or 2.  The measurement of the level lifetime is extracted from the ground-state $\gamma$ ray.

2622.1 keV $(2^+)$ level.  The level is established based on the observation of a $\gamma$ ray to the ground state with a 2.9 MeV threshold.  The $a_2$ coefficient is small, yet positive, thus a spin-parity of $(2^+)$ is tentatively assigned.

2694.0 keV $3^-$ state.  Only a 2026.2 keV $\gamma$ ray to the $2^+_1$ state is observed.  From the angular distribution comparison with \textsc{cindy} \cite{cindy} calculations, the spin is 3 and is a pure $E1$ transition, indicating negative parity.

2721.5 keV $2^+$ state.  While the ground-state $\gamma$ ray is mixed with background, it is definitively present with a threshold of 2.8 MeV from the excitation function.  The background is isotropic (based on measurements for other nuclei), and the angular distribution is quadrupole, indicating a spin-parity of $2^+$.

2758.1 keV $2^+$ state.  A 2758.1 keV $\gamma$ ray is observed with a threshold of 2.8 MeV and a quadrupole angular distribution, establishing a $2^+$ level at this energy.

2781.3 keV level.  Gamma rays representing transitions to the $2^+_1$ and $2^+_2$ states are observed with 2.9 MeV thresholds.  No information concerning the spin of the level could be extracted, however.

2818.5 keV $(3^-)$ level.  A 2150.8 keV $\gamma$ ray was observed at a threshold of 2.9 MeV, with an angular distribution that most closely compares with the \textsc{cindy}  \cite{cindy} calculations for spin 3 with no mixing ratio, thus indicating negative parity.

2821.1 keV $1,2^+$ level.  Only a $\gamma$ ray to the ground state is observed from this level with an ill-defined angular distribution, limiting the spin and parity to 1 or $2^+$.

2896.6 keV $(3^+)$ level.  Beginning at an incident neutron energy of 3.0 MeV, $\gamma$ rays at 1598.7 and 2228.9 keV are observed.  The angular distribution of the 1598.7 keV $\gamma$ ray when compare best with \textsc{cindy} \cite{cindy} calculations for spin 3 and has a mixing ratio, thus we conclude a tentative spin-parity of $(3^+)$ for the level.  The angular distribution results for the much weaker 2228.9 keV $\gamma$ ray are inconclusive.

2903.0 keV $2^+,4^+$ level.  Gamma rays at 1099.1 and 2235.4 keV are observed at a threshold of 3.0 MeV and placed as transitions to the $3^+_1$ and $2^+_1$ states, respectively.  The angular distributions when compared with \textsc{cindy} \cite{cindy} calculations allow us to limit the spin to either $2^+$ or $4^+$, but a ground-state $\gamma$ ray is not observed.

2922.3 keV 1 level.  A single 2922.3 keV $\gamma$ ray is observed with a dipole angular distribution, indicating a ground-state transition from a spin 1 state.

2928.9 keV $2^+$ state.  Gamma rays representing transitions to the $2^+_1$ state and the ground state are observed.  The angular distribution of the 2928.9 keV $\gamma$ ray is quadrupole in shape, allowing the conclusion that the state has a $2^+$ spin and parity.

2960.0 keV $1,2^+$ level.  Only a $\gamma$ ray to the ground state is observed with a nondescript angular distribution, limiting the spin and parity to 1 or $2^+$.

2969.0 keV $1,2^+$ level.  Gamma rays to the $2^+_2$, $2^+_1$, and ground states are observed, but the spin and parity of this new level can only be limited to 1 or $2^+$.

3050.8 keV $2^+$ state.  A ground-state $\gamma$ ray with a quadrupole angular distribution defines the spin-parity of the level as $2^+$.

3091.6 keV 1 level.  A ground-state $\gamma$ ray with a dipole angular distribution defines the spin of the level to be 1.  Although depopulating $\gamma$ rays to the $2^+_2$ and $2^+_1$ states are also observed, the parity could not be determined from these weaker branches.

3113.2 keV $2^+$ state.  A ground-state $\gamma$ ray with a quadrupole angular distribution defines the spin-parity of the level as $2^+$.

3145.5 keV level.  A 2477.8 keV $\gamma$ ray is placed as a transition to the $2^+_1$ state based on its 3.4 MeV threshold, but no information concerning the spin of this weakly populated level could be obtained.

3274.4 keV level.  Based on its 3.4 MeV threshold, a 2606.7 keV $\gamma$ ray is placed as a transition to the $2^+_1$ state.  Again, information concerning the spin of this weakly populated level could not be obtained.

\subsection{Other levels of interest}

%
%

2111.9 keV $6^+$ state.  The history of this level is quite complex, and from our work, we refute its existence.  Hamilton \textit{et al.} \cite{Ham65} originally proposed this level based on the observation of $\gamma$-ray doublets at 669, 671 keV, and 727, 729 keV, where the former pair was thought to depopulate states at 2110 and 2112 keV (decaying to the 1440 keV $4^+_1$ state), and the latter pair to feed those levels from a 2839 keV level.  However, a subsequent publication \cite{Car70} states that the 729 keV $\gamma$ ray arose only from an impurity.  Yet another publication by Hamilton \textit{et al.} \cite{Ham70} re-establishes the 2110 and 2112 keV levels based on coincidence data using one Ge(Li) detector and one NaI detector, still believing the 669-671 keV doublet exists.  Kerek \textit{et al.} \cite{Ker71} using data from the $(\alpha,2n\gamma)$ reaction proposed that the 727 keV $\gamma$ ray feeds the 1440 keV state directly, eliminating the 2839 keV level and establishing a 2167 keV level.  Still further confusion arises when Singhal \textit{et al.} \cite{Sin73} claim the 729 keV $\gamma$ ray is not entirely an impurity, and Girit \textit{et al.} \cite{Gir80} question the existence of the 669-671 keV doublet, but still conclude a spin-parity of $6^+$ for the 2112 keV level.  In our INS measurements, we find no evidence of a 671 keV $\gamma$ ray, nor a 729 keV $\gamma$ ray; we refute the existence of the doublets and, therefore, the existence of the 2112 keV level.  Recent results reported by Vogt \textit{et al.} \cite{Vogt.PhysRevC.96.024321} from measurements employing multinucleon-transfer and fusion-evaporation reactions do not include the observation of decaying or feeding transitions associated with the 2112 keV level either.

2167.4 keV $6^+$ state.  As noted previously in the discussion of the 2111.9 keV state, this level was proposed by Kerek \textit{et al.} \cite{Ker71} from $(\alpha,2n\gamma)$ measurements.  The observed threshold for the 727 keV $\gamma$ ray in our measurements is 2.2 MeV, in agreement with its placement as directly feeding the 1440 keV state.  There is no clear indication that the 727 keV $\gamma$ ray is a doublet for $ E_n \ge$ 2.9 MeV.  Kerek \textit{et al.} \cite{Ker71} favored a $5^+$ spin-parity assignment, which Girit \textit{et al.} \cite{Gir80} supported based on the angular correlation of the 417 feeding $\gamma$ ray in $\beta$-decay measurements.  Although it would be mixed with background, no 417 keV feeding $\gamma$ ray is observed in our measurements based on the comparison of the intensity of the 417 keV $\gamma$ ray in spectra for other nuclei.  Vogt \textit{et al.} \cite{Vogt.PhysRevC.96.024321} obtained angular correlation data for the 727 keV $\gamma$ ray as well, and used it as a benchmark for their measurements with a fit for a $5^+ \rightarrow 4^+ \rightarrow 2^+$ cascade.  However, from our angular distribution data for the 727 keV $\gamma$ ray, we rather assign a $6^+$ spin-parity.

2169.3 keV $0^+_3$ state.  As described in Ref. \cite{Pet16}, this state was previously assigned as $J^\pi=1,2^+$ in Ref. \cite{NDS132}, but we establish a spin-parity of $0^+$.

2512.0 keV $4^-$ state.  The NDS compilation \cite{NDS132} assigns a spin-parity of $(4^+)$ for this level, presumably based upon having decays to both the $5^-_1$ and $2^+_4$ states.  In our work, however, we do not observe the 325 keV $\gamma$ ray to the $2^+_4$ state, and based on the angular distribution and excitation function data, we prefer a $4^-$ assignment.  Hamada \textit{et al.} \cite{Ham88}, also assigned the spin and parity as $4^-$.

2555.7 keV $3^-$ state.  The NDS compilation \cite{NDS132} lists a spin-parity of $(2^+,3)$ for this level.  In our data, the angular distributions of both the 570.0 and 1888.0 keV $\gamma$ rays when compared with \textsc{cindy} \cite{cindy} calculations indicate spin 3 and pure $E1$ multipolarity, thus we conclude $J^\pi=3^-$.

2839.4 and 2840.2 keV levels.  These levels are separated on the basis of $\gamma$-ray energies only.  No spin information could be obtained from any of the assigned $\gamma$ rays.

2872.7 keV $(4,6)^-$ level.  No prior spin assignment for this level has been given \cite{NDS132}, but we conclude it is $(4,6)^-$ based on the angular distribution of the 832.3 keV $\gamma$ ray and comparisons with \textsc{cindy} \cite{cindy} calculations.




\section{Shell-Model Calculations}

As noted previously, shell-model calculations have been recently reported for the Xe isotopes by Teruya \textit{et al.} \cite{Teruya.PhysRevC.92.034320}, Van Isacker \textit{et al.} \cite{Xe136.PhysRevC.98.034302}, and Vogt \textit{et al.} \cite{Vogt.PhysRevC.96.024321,Vogt.PhysRevC.93.054325}.

Teruya \textit{et al.} \cite{Teruya.PhysRevC.92.034320}, who performed shell-model calculations for even-even, odd-mass, and odd-odd nuclei of Sn, Sb, Te, I, Xe, Cs, and Ba isotopes around mass 130, used a phenomenological effective interaction based on an extended pairing plus quadrupole-quadrupole interaction. They also truncated the model space by first diagonalizing the Hamiltonian separately for protons and neutrons to select the most important configurations. The numbers of states in the proton and neutron model spaces were then increased until convergence was reached.

Van Isacker \cite{Xe136.PhysRevC.98.034302} used the N82K interaction, an empirical interaction derived for $N=82$ nuclei by Kruse and Wildenthal \cite{N82K}.  Vogt \textit{et al.} \cite{Vogt.PhysRevC.96.024321,Vogt.PhysRevC.93.054325} used interactions derived by Brown \textit{et al.} \cite{BRN.PhysRevC.71.044317} based on the CD Bonn nucleon-nucleon interaction, which have also proven successful in describing the electromagnetic properties of low-lying states around $^{132}$Sn \cite{AES.Te134.PhysRevC.88.051304,AES-Te136RIV.PhysRevC.96.014321}. These same interactions, designated jj55 (or sn100pn), are employed in the calculations reported here.


Shell-model calculations were performed with the \textsc{NuShellX@MSU} code \cite{NuShellX} for the isotopes $^{136}$Xe, $^{134}$Xe, and $^{132}$Xe having four protons and zero, two, and four neutron holes, respectively, relative to $^{132}$Sn. All proton and neutron single-particle orbitals in the $50-82$ shell ($ 0g_{7/2}, 1d_{5/2}, 1d_{3/2}, 2s_{1/2}, 0h_{11/2}$) were included. Single-particle energies were set by reference to the low-lying states of  $^{133}$Sb and $^{131}$Sn for proton particles and neutron holes, respectively. As described in Refs.~\cite{BRN.PhysRevC.71.044317,JMATe2.PhysRevC.90.014322,JMATe136}, the interactions are based on the CD Bonn potential with the renormalization of the $G$ matrix carried to third order. A Coulomb term is added to the proton-proton interaction.

As in Ref.~\cite{AES-Te136RIV.PhysRevC.96.014321}, the effective $M1$ operator applied a correction $\delta g_l(p) = 0.13$ to the proton orbital $g$~factor and quenched the spin $g$~factors for both protons and neutrons to $70\%$ of their bare values. (The tensor term was ignored.) The effective $M1$ operator is similar to that of Jakob \textit{et al.}~\cite{Xepaper.PhysRevC.65.024316} and in reasonable agreement with that of Brown \textit{et al.}~\cite{BRN.PhysRevC.71.044317}.

Calculations with the same basis, interactions, and $M1$ operator were reported in Ref.~\cite{AES-Te136RIV.PhysRevC.96.014321} for the $N=78$, 80, and 82 isotopes of Te and Xe, with an emphasis on $g$~factors and $B(E2; 2^+_1\rightarrow 0^+_1)$  values. Overall, the description of these electromagnetic observables was good, although there remained some shortfall in $E2$ strength when the effective charges were set to the standard values of $e_p = 1.5e$ and $e_n=0.5e$.  We, therefore, began by studying the nuclei $^{136}$Xe, $^{130}$Sn, and $^{128}$Sn, in order to set the effective charges, and also gain insight into the proton and neutron structures that combine to form the states in $^{132}$Xe and $^{134}$Xe.


\subsection{$^{136}$Xe: Proton configurations and the proton effective charge}

The $B(E2)$ values are related to the effective charges of the proton ($e_p$) and neutron ($e_n$) by
\begin{equation}
B(E2;J_i \rightarrow J_f) = \frac{1}{(2J_i+1)} \left [e_p A_p + e_n A_n \right ]^2,
\end{equation}
where \textsc{NushellX} reports the values of $A_p$ and $A_n$ and the effective charges are in units of the elementary charge $e$. As $A_n=0$ for the $N=82$ nuclide $^{136}$Xe, a comparison of measured and calculated $B(E2; 2^+_1 \rightarrow 0^+_1)$ values determines the proton effective charge. The adopted experimental $B(E2)$ is a weighted average of the  values reported in Refs.\cite{Xepaper.PhysRevC.65.024316,Stahl-thesis}, namely $B(E2; 2^+_1 \rightarrow 0^+_1)=10.0(3)$~W.u., which yields $e_p=1.74(3)$.  The error reflects the uncertainty in the experimental $B(E2)$. This effective charge is essentially the same as that used in Ref. \cite{Xe136.PhysRevC.98.034302}, $e_p=1.73$. Teruya \textit{et al.} \cite{Teruya.PhysRevC.92.034320} have a $Z$-dependent effective charge, which takes the slightly smaller value of $e_p=1.6$ for the Xe isotopes. Given the uncertainties in evaluating the effective charge, $e_p=1.7$ is adopted for the following calculations. The effective charge will be discussed again at the end of this section, after considering the excited-state wavefunctions.

Previous work on $^{136}$Xe reported shell-model calculations in the same basis as those reported here, but with alternative interactions. The present and previous calculations of $B(E2)$ transition rates are compared in Table~\ref{tab:136XeE2}. Overall, there is reasonable agreement between the calculations and experiment, with the N82K interactions \cite{Xe136.PhysRevC.98.034302} generally giving the better description. The calculations of Teruya \textit{et al.} \cite{Teruya.PhysRevC.92.034320} consistently fall below the N82K calculations \cite{Xe136.PhysRevC.98.034302}, due, at least in part, to their use of a smaller proton charge. Differences between the N82K \cite{Xe136.PhysRevC.98.034302} and the present jj55 calculations stem from differences in the wavefunctions, which are compared for these two interactions in Appendix \ref{ApA}.

\begin{table}[]
\centering
\caption{$E2$ transition rates in $^{136}$Xe.}
\label{tab:136XeE2}
\begin{ruledtabular}
\begin{tabular}{lcccc}
Transition  & \multicolumn{4}{c}{$B(E2)$ ($e^2$fm$^4$)}   \\ \cline{2-5}
           & \multicolumn{1}{c}{Exp. \cite{NDS136,Xe136.PhysRevC.98.034302}} & \multicolumn{1}{c}{Ref.~\cite{Teruya.PhysRevC.92.034320}} & \multicolumn{1}{c}{N82K \cite{Xe136.PhysRevC.98.034302}} & \multicolumn{1}{c}{jj55} \\
\hline
\\
$2^+_1 \rightarrow 0^+_1$       &   415(12)         & 357   & 400   & 398 \\
$4^+_1 \rightarrow 2^+_1$       &   53.2(7)         & 63.6  &  86   &  48 \\
$6^+_1 \rightarrow 4^+_1$       &   0.55(2)         & 0.088 & 0.12  &   4.8 \\
$2^+_2 \rightarrow 0^+_1$       &   23(3)           &       &  12   &  48 \\
$2^+_3 \rightarrow 0^+_1$       &   38(3)           &       &  12   &  0.7 \\
$2^+_4 \rightarrow 0^+_1$       &   12(6)           &       &  9.6  &  24 \\
$2^+_5 \rightarrow 0^+_1$       &   21.7(15)        &       &  22   &  4 \\
$2^+_2 \rightarrow 2^+_1$       &   299(71)         &       & 103   &  8 \\
$2^+_3 \rightarrow 2^+_1$       &  21$^{+58}_{-21}$ &       & 117   & 308 \\
\end{tabular}
\end{ruledtabular}
\end{table}

The two interactions generally predict wavefunctions with the same dominant components, often with similar amplitudes (see Appendix \ref{ApA}). The main difference, evident in both the $E2$ transition strengths and from inspection of the wavefunctions, concerns the character of the 2$^+_2$ and 2$^+_3$ states, which are approximately interchanged in character in the jj55 calculations compared with the N82K calculations \cite{Xe136.PhysRevC.98.034302}. The jj55 interaction has the 2$^+_3$ state as being the (predominantly) seniority $\upsilon=4$, $\pi (g_{7/2}^4)_{2^+}$ state, which explains the small predicted transition strength to the ground state. This assignment is not supported by the experimental $B(E2; 2^+ \rightarrow 0_1^+)$ values, which indicate mixed states, favoring a larger $\upsilon=4$, $\pi (g_{7/2}^4)_{2^+}$ contribution in the 2$^+_2$ state. The character of the 2$^+_4$ and 2$^+_5$ states also appears to be interchanged in the jj55 calculations.
It is not unexpected that the empirically derived N82K interaction, tuned to $N=82$, is able to explain some of the finer details with greater accuracy than the jj55 interaction, which is derived from a nucleon-nucleon interaction based on effective field theory.

Previous work on $^{136}$Xe reported schematic calculations in a limited basis of $\pi g_{7/2}^m \otimes d_{5/2}^n$ where $m+n=4$, and the cases of $n=0,1,2$ account for all of the states below about 2.8 MeV \cite{Xe136.PhysRevC.98.034302}. The large-basis calculations (see Appendix \ref{ApA}) support the proposed dominant configurations up to the 4$^+_2$ state, but there is considerable configuration mixing. Above the 4$^+_2$ state, there are strong variations from the simple picture.

The procedure adopted to set the proton effective charge requires further discussion. First, the transition rates for the $4^+_1 \rightarrow 2^+_1$ and  $6^+_1 \rightarrow 4^+_1$ transitions, which are known experimentally, do not provide a reliable means to set the effective charge. The reason is that $E2$ transitions between these members of the $\upsilon=2$ $\pi g_{7/2}^4$ configuration are forbidden by the so-called midshell cancellation (the $\pi g_{7/2}$ orbital is half full) \cite{midshell}. Thus these $B(E2)$ values arise entirely from configuration mixing; they are, therefore, very sensitive to the degree of configuration mixing and cannot give a clear indication of the proton effective charge. Second, some discussion of the effects of quadrupole collectivity in the $^{132}$Sn core is in order. For this purpose, we refer to the case of $^{134}$Te. In this nuclide, the $g$~factors of the 2$^+$, 4$^+$, and 6$^+$ members of the nominal $\pi g_{7/2}^2$ configuration, as well as the $E2$ transition rates for the decays of these states, have all been measured \cite{AES.Te134.PhysRevC.88.051304}. There is evidence of extra collectivity in the 2$^+$ state, which can be described by including a small admixture due to 2$^+$ excitations of the $^{132}$Sn core (i.e., particle-vibration coupling). However, the overall conclusion from comparing the experimental and theoretical moments and $E2$ transition rates was that the contribution of core excitation is modest, and that $^{132}$Sn is a good doubly magic nucleus. Setting the effective charge empirically, as done here, implicitly includes contributions from coupling of the first excited state to the quadrupole excitations of the core, but this contribution can be expected to be small.


\subsection{$^{128,130}$Sn: Neutron configurations and the neutron effective charge}

Table~\ref{tab:SM130Sn} shows results of shell-model calculations for $^{130}$Sn based on the jj55 interaction. The theoretical excitation energies in $^{130}$Sn agree quite well with experiment (where data are available). These states provide a reference for the neutron configurations in the isotone $^{134}$Xe. A feature of the neutron space near $N=82$ is that the $2s_{1/2}$, $1d_{3/2}$, and $0h_{11/2}$ orbitals are all close in energy. Thus, strongly mixed neutron wavefunctions are expected. Nevertheless, the two-neutron hole states in $^{130}$Sn at low excitation energies are dominated by the $\nu h_{11/2}^{-2}$ configuration. This configuration becomes more prominent in the yrast states as the spin increases, and is unique for the 8$^+$ and 10$^+$ states.

Unfortunately, the $B(E2; 2^+_1 \rightarrow 0^+_1)$ for $^{130}$Sn is not known, so it cannot be used to estimate the neutron effective charge. However, the lifetime of the (10$^+$) state is known giving the $B(E2; 10^+_1 \rightarrow 8^+_1) = 0.38(4)$~W.u. =14.9(1.6) $e^2$fm$^4$ \cite{NDS.A130.SINGH200133}. Within the jj55 model space, this transition is a pure
$\nu (h_{11/2}^{-2})_{10^+} \rightarrow \nu (h_{11/2}^{-2})_{8^+}$ transition. The observed $B(E2)$ implies $e_n = 0.838(45)$. This value agrees with $e_n = 0.8$ as adopted by Teruya \textit{et al.} \cite{Teruya.PhysRevC.92.034320} for $N=130$.

\begin{table*}[]
\centering
\caption{Shell-model results for $^{130}$Sn using the jj55 interaction.}
\label{tab:SM130Sn}
\begin{ruledtabular}
\begin{tabular}{crrl}
$J^{\pi}_i$  & \multicolumn{2}{c}{$E_x$ (MeV)}  & Wavefunction \\ \cline{2-3}
           & Exp. \cite{NDS.A130.SINGH200133}& Theory &\\
\hline
0$^+_1$  & 0      &  0        &  $51.5\% (h_{11/2}^{-2}) + 25.1\% (d_{3/2}^{-2}) + 9.1\% (s_{1/2}^{-2}) + 8.6\% (d_{5/2}^{-2}) + ...$   \\
2$^+_1$  & 1.221  & 1.380  &  $56.4\% (h_{11/2}^{-2}) + 19.4\% (s_{1/2}^{-1}d_{3/2}^{-1}) + 11.3\% (d_{3/2}^{-2}) + ...$ \\
0$^+_2$  &        & 1.920  &  $43.0\% (d_{3/2}^{-2}) + 40.4\% (h_{11/2}^{-2}) + 16.0\% (s_{1/2}^{-2}) + ...$   \\
4$^+_1$  & 1.996  & 2.077  &  $95.5\% (h_{11/2}^{-2}) + ...$  \\
2$^+_2$  & 2.028  & 2.003  &  $39.6\% (d_{3/2}^{-2}) + 37.7\% (h_{11/2}^{-2}) + 17.3\% (s_{1/2}^{-1}d_{3/2}^{-1}) + ...$  \\
6$^+_1$  & 2.257  & 2.278  &  $99.6\% (h_{11/2}^{-2}) + ...$  \\
0$^+_3$  &        & 2.339  &  $74.3\% (s_{1/2}^{-2}) + 25.4\% (d_{3/2}^{-2}) + ...$   \\
8$^+_1$  & 2.338  & 2.357  &  $100.0\% (h_{11/2}^{-2})$  \\
10$^+_1$ & 2.435  & 2.418  &  $100.0\% (h_{11/2}^{-2})$  \\

\end{tabular}
\end{ruledtabular}
\end{table*}

\begin{table*}[]
\centering
\caption{Shell-model results for $^{128}$Sn using the jj55 interaction.}
\label{tab:SM128Sn}
\begin{ruledtabular}
\begin{tabular}{crrl}
$J^{\pi}_i$  & \multicolumn{2}{c}{$E_x$ (MeV)}  & Wavefunction \\ \cline{2-3}
           & Exp. \cite{ELEKES2015191} & Theory &\\
\hline
0$^+_1$  & 0      &  0        &  $34.1\% (h_{11/2}^{-2} d_{3/2}^{-2}) + 19.8\% (h_{11/2}^{-4}) + 10.4\% (h_{11/2}^{-2} s_{1/2}^{-2}) + 8.7\% + (h_{11/2}^{-2} d_{5/2}^{-2}) + 7.3\% (h_{11/2}^{-2} g_{7/2}^{-2}) + ...$    \\

2$^+_1$  & 1.169  &  1.197  &  $35.9\% (h_{11/2}^{-2} d_{3/2}^{-2}) + 13.6\% (h_{11/2}^{-4}) + 12.7\% (h_{11/2}^{-2} s_{1/2}^{-1} d_{3/2}^{-1}) + 6.5\% (h_{11/2}^{-2} s_{1/2}^{-2}) + 5.4\% (h_{11/2}^{-2} d_{5/2}^{-2}) + ...$ \\

4$^+_1$  & 2.000  &  1.977  &  $44.3\% (h_{11/2}^{-2} d_{3/2}^{-2}) + 18.4\% (h_{11/2}^{-4}) + 9.9\% (h_{11/2}^{-2} s_{1/2}^{-2}) + 6.8\% (h_{11/2}^{-2} d_{5/2}^{-2}) + 5.2\% (h_{11/2}^{-2} g_{7/2}^{-2}) + ...$   \\

2$^+_2$  & 2.104  &  1.979  &  $35.3\% (h_{11/2}^{-2} d_{3/2}^{-2}) + 33.2\% (h_{11/2}^{-2} s_{1/2}^{-1} d_{3/2}^{-1}) + 5.2\% (h_{11/2}^{-2} s_{1/2}^{-1} d_{5/2}^{-1}) + 5.0\% (h_{11/2}^{-2} d_{3/2}^{-1}g_{7/2}^{-1}) + ...$   \\

0$^+_2$  &        &  2.159  &  $47.4\% (h_{11/2}^{-4}) + 19.5\% (s_{1/2}^{-2} d_{3/2}^{-2}) + 5.4\% (h_{11/2}^{-2} s_{1/2}^{-2}) + 5.3\% (h_{11/2}^{-2} g_{7/2}^{-2}) + ...$  \\

6$^+_1$  &        &  2.271  &  $44.8\% (h_{11/2}^{-2} d_{3/2}^{-2}) + 23.3\% (h_{11/2}^{-4}) + 11.4\% (h_{11/2}^{-2} s_{1/2}^{-2}) + 7.6\% (h_{11/2}^{-2} d_{5/2}^{-2}) + 5.7\% (h_{11/2}^{-2} g_{7/2}^{-2}) + ...$  \\
0$^+_3$  &        &  2.330  &  $37.3\% (h_{11/2}^{-2} s_{1/2}^{-2}) + 32.2\% (h_{11/2}^{-2} d_{3/2}^{-2})   + 8.5\% (s_{1/2}^{-2} d_{3/2}^{-2}) + ...$   \\

2$^+_3$  & 2.258  &  2.335  &  $52.7\% (h_{11/2}^{-2} d_{3/2}^{-2}) + 35.9\% (h_{11/2}^{-2} s_{1/2}^{-1} d_{3/2}^{-1}) + 11.7\% (h_{11/2}^{-4}) + ...$    \\

8$^+_1$  & 2.413  &  2.377  &  $43.8\% (h_{11/2}^{-2} d_{3/2}^{-2}) + 24.8\% (h_{11/2}^{-4}) + 11.5\% (h_{11/2}^{-2} s_{1/2}^{-2}) + 7.8\% (h_{11/2}^{-2} d_{5/2}^{-2}) + 5.8\% (h_{11/2}^{-2} g_{7/2}^{-2}) + ...$  \\
10$^+_1$ & 2.492  &  2.410  &  $45.3\% (h_{11/2}^{-2} d_{3/2}^{-2}) + 24.6\% (h_{11/2}^{-4}) + 11.7\% (h_{11/2}^{-2} s_{1/2}^{-2}) + 7.8\% (h_{11/2}^{-2} d_{5/2}^{-2}) + 5.9\% (h_{11/2}^{-2} g_{7/2}^{-2}) + ...$  \\

\end{tabular}
\end{ruledtabular}
\end{table*}

Table~\ref{tab:SM128Sn} shows results of the shell-model calculations for $^{128}$Sn, which provide a reference for the neutron configurations in the isotone $^{132}$Xe. The agreement between the theoretical and experimental excitation energies is good. Given the dominance of the $\nu h_{11/2}^{-2}$ configuration in $^{130}$Sn, it might have been expected that $\nu h_{11/2}^{-4}$ would dominate in $^{128}$Sn; however, this is not the case. Instead the prominent configuration in the low-lying states of $^{128}$Sn is $\nu h_{11/2}^{-2}d_{3/2}^{-2}$.

The $B(E2; 0^+_1 \rightarrow 2^+_1)$ for $^{128}$Sn has been measured by Allmond \textit{et al.} \cite{JMATe136}.  The effective charge implied is  $e_n = 0.80(3)$.
A similar analysis on $^{126}$Sn gives $e_n = 0.83(3)$. Thus these effective charges deduced from $^{126}$Sn, $^{128}$Sn and $^{130}$Sn are all consistent with $e_n = 0.8$. The lifetime of the (10$^+$) state in $^{128}$Sn is also known giving the $B(E2; 10^+_1 \rightarrow 8^+_1) = 0.346(18)$~W.u. =13.3(7) $e^2$fm$^4$ \cite{ELEKES2015191}, which implies a smaller effective charge of $e_n = 0.46(1)$. The wavefunctions of the 8$^+_1$ and 10$^+_1$ states in $^{128}$Sn are more complex than in $^{130}$Sn so the theoretical uncertainty might exceed the experimental uncertainty quoted for this effective charge. Nevertheless, we do not find evidence for an increased effective charge ($e_n = 1$) for $^{128}$Sn as used by Teruya \textit{et al.} \cite{Teruya.PhysRevC.92.034320}. Given our focus on the low-lying states, $e_n = 0.8$ is adopted for the following calculations.

\subsection{Results}

\subsubsection{$^{134}$Xe}

\begin{figure*}[ht!]
\centering
\includegraphics[width = 13 cm, keepaspectratio, trim={0 0 0 0cm}, clip]{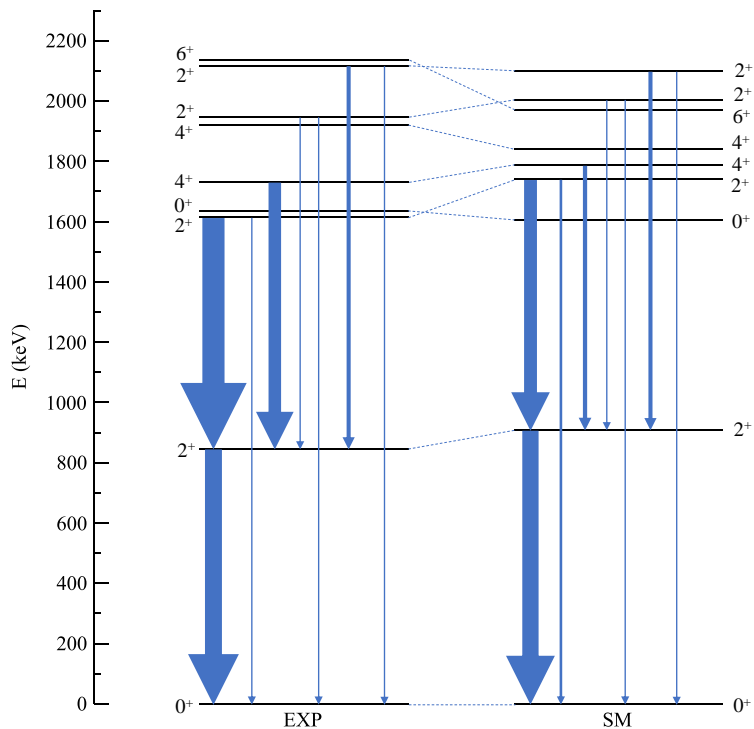}
\caption{Comparison of the experimental results with those of the present shell-model calculations using the jj55 interaction for $^{134}$Xe.  The widths of the arrows are proportional to the $B(E2)$ values.  See Table~\ref{tab:SM_E2_134Xe} for the experimental values with uncertainties.}~\label{fig:134SM}
\end{figure*}

Experimental and theoretical excitation energies and $E2$ transition rates for $^{134}$Xe are compared in Table~\ref{tab:SM_Ex_134Xe} ($E_x$), Table~\ref{tab:SM_E2_134Xe} ($B(E2)$) and Fig.~\ref{fig:134SM}. 
The dominant components of the wavefunctions are listed for selected low-spin states up to $6_1^+$  in Appendix \ref{ApB}.

Overall, the agreement between experimental and theoretical level energies is good. The vibrator-like level sequence is reproduced by the shell model.  However, apart from the $B(E2)$ for the  $2^+_1 \rightarrow 0^+_1$ transition, which is well described, the $E2$ transition rates tend to be underestimated. The calculations of Teruya \textit{et al.} \cite{Teruya.PhysRevC.92.034320} also reproduce $B(E2; 2^+_1 \rightarrow 0^+_1)$, but tend to over-estimate the transition strengths between higher-lying excited states. They use the same effective charges as we do for $^{134}$Xe, so differences must stem from the wavefunctions. Note that they report calculations only for selected transitions.

There are two predicted states near the observed 6$^+$ state and likewise two predicted 8$^+$ states near the observed 8$^+$ state. The electromagnetic decay properties may suggest that the observed 6$^+$ state is closer to the wavefunction of the second one predicted (which actually has an energy closer to the observed state), and likewise for the 8$^+$ states. More detailed spectroscopy using a reaction such as heavy-ion Coulomb excitation is needed to confirm the yrast nature of the observed states and find the predicted nearby yrare 6$^+$ and 8$^+$ states.

Although there is not good quantitative agreement on the $E2$ transition strengths, most patterns are correct. For example, the present shell-model calculations correctly predict that the $2^+_i \rightarrow 2^+_1$ transitions are much stronger than the $2^+_i \rightarrow 0^+_1$ transitions, where~$i \geq 2$.

The spin decomposition of the wavefunctions is given along with a comparison of experimental and theoretical $g$~factors in Table~\ref{tab:spincomp134Xe}. The present theoretical $g$~factors are in very good agreement with experiment. Those of Teruya \textit{et al.} \cite{Teruya.PhysRevC.92.034320} are uniformly smaller than the present calculations and underestimate the experimental values. This difference most likely stems from their choice of the effective orbital $g$~factor for protons in the $M1$ operator. As described in previous work \cite{AES.Te134.PhysRevC.88.051304,AES-Te136RIV.PhysRevC.96.014321}, we adopt $g_\ell^{\pi} = 1.13$, rather than the bare value of $g_\ell^{\pi} = 1$. This relatively small change in $g_\ell^{\pi} $ is amplified in the $g$~factors of the low-lying states of the Xe isotopes because the $M1$ operator depends on $g_{\ell} \ell$, where $\ell$ is the orbital angular momentum, and the $\pi g_{7/2}$ orbital with $\ell = 4$ is prominent in the configurations of the low-lying states.

\begin{table}[]
\centering
\caption{Excitation energies in $^{134}$Xe using the jj55 interaction.}
\label{tab:SM_Ex_134Xe}
\begin{ruledtabular}
\begin{tabular}{crl}
$J^{\pi}_i$  & \multicolumn{2}{c}{$E_x$ (MeV)}  \\ \cline{2-3}
  & Exp. \cite{NDS134,Pet17B}  & SM \\
\hline
\\
0$^+_1$  &  & 0       \\
2$^+_1$  &  0.847 & 0.909   \\
2$^+_2$  &  1.614 & 1.740   \\
0$^+_2$  &  1.636 & 1.607   \\
4$^+_1$  &  1.731 & 1.788    \\
4$^+_2$  &  1.920 & 1.841   \\
2$^+_3$  &  1.947 & 2.004   \\
2$^+_4$  &  2.117 & 2.099   \\
6$^+_1$  &  2.137 & 1.971   \\
6$^+_2$  &        & 2.081   \\
8$^+_1$  &  2.997 & 2.920   \\
8$^+_2$  &        & 2.976   \\

\\
\end{tabular}
\end{ruledtabular}
\end{table}

\begin{table}[]
\centering
\caption{Transition rates in $^{134}$Xe.}
\label{tab:SM_E2_134Xe}
\begin{ruledtabular}
\begin{tabular}{ccccc}
Transition  & $B(E2)_{\rm exp}$  & \multicolumn{3}{c}{$B(E2)$ ($e^2$fm$^4$)}  \\ \cline{3-5}
 & (W.u.) & Exp. \cite{NDS134,Pet17B} & SM Ref.~\cite{Teruya.PhysRevC.92.034320} & Present SM\\
\hline
\\
  $2^+_1 \rightarrow 0^+_1$    & 15.3(11)           &   623(45)           &  623    & 601  \\

  $2^+_2 \rightarrow 2^+_1$    & 20(2)              &   815(81)           &         & 476  \\

  $2^+_2 \rightarrow 0^+_1$    & 0.72$^{+19}_{-18}$  &   31$^{+8}_{-7}$    &         &  94  \\

  $0^+_2 \rightarrow 2^+_1$    & $<55$              &   $<2240$           &         &  44  \\

  $4^+_1 \rightarrow 2^+_1$    & 11.6(8)            &   460(33)           &   758   & 157  \\

  $4^+_2 \rightarrow 2^+_1$    & $<16$              &   $<650$            &         & 557  \\

  $2^+_3 \rightarrow 2^+_2$    & 0.26$^{+56}_{-24}$  & 10.6$^{+23}_{-10}$  &         &  19  \\
\multicolumn{1}{r}{or}     & 75.6$^{+76}_{-75}$  &   3080(310)         &         &   \\
  $2^+_3 \rightarrow 0^+_1$    & 0.755$^{+81}_{-76}$ &    31(3)            &         & 2  \\

  $2^+_4 \rightarrow 2^+_2$    & 3.6$^{+16}_{-12}$   & 147$^{+65}_{-49}$   &         &  149  \\
\multicolumn{1}{r}{or}       & 6.3(18)            &   260(70)           &         &   \\
  $2^+_4 \rightarrow 0^+_1$    & 0.056$^{+16}_{-14}$ &    2.3$^{+7}_{-6}$  &         & 17  \\

  $6^+_1 \rightarrow 4^+_1$    &                    &                     &   140   & 7  \\

  $6^+_2 \rightarrow 4^+_1$    &                    &                     &         & 123  \\

  $8^+_1 \rightarrow 6^+_1$    &                    &                     &   202   &  465 \\

  $8^+_1 \rightarrow 6^+_2$    &                    &                     &      &  14 \\

  $8^+_2 \rightarrow 6^+_1$    &                    &                     &   202   & 5  \\

  $8^+_2 \rightarrow 6^+_2$    &                    &                     &      &   541 \\

\\
\end{tabular}
\end{ruledtabular}
\end{table}

\begin{table*}[]
\centering
\caption{Spin decomposition and $g$~factors in $^{134}$Xe.}
\label{tab:spincomp134Xe}
\begin{ruledtabular}
\begin{tabular}{cllll}
$J^{\pi}_i$  & \multicolumn{3}{c}{$g$}  & Spin composition \\ \cline{2-4}
            & Exp. \cite{Xepaper.PhysRevC.65.024316} & SM Ref.~\cite{Teruya.PhysRevC.92.034320} & Present SM\\
\hline
\\
2$^+_1$  & 0.354(7)  & 0.324  & 0.411  &  $ 0.46 \pi(0^+) \nu(2^+) + 0.39 \pi(2^+) \nu(0^+) + ...$ \\
2$^+_2$  &           &        & 0.531  &  $ 0.39 \pi(2^+) \nu(0^+) + 0.31 \pi(0^+) \nu(2^+) + 0.20 \pi(2^+) \nu(2^+) + ...$  \\
4$^+_1$  & 0.83(14)  & 0.555  & 0.855  &  $ 0.78 \pi(4^+) \nu(0^+)  + ...$   \\
4$^+_2$  &           &        & 0.619  &  $ 0.45 \pi(4^+) \nu(0^+) + 0.26 \pi(2^+) \nu(2^+) + 0.11 \pi(0^+) \nu(4^+) + ...$   \\
6$^+_1$  &           & 0.690  & 0.934  &  $ 0.78 \pi(6^+) \nu(0^+)  + ...$  \\
6$^+_2$  &           &        & 1.498  &  $ 0.75 \pi(6^+) \nu(0^+)  + ...$  \\
\end{tabular}
\end{ruledtabular}
\end{table*}

\subsubsection{$^{132}$Xe}

\begin{figure*}[ht!]
\centering
\includegraphics[width = 13 cm, keepaspectratio, trim={0 0 0 0cm}, clip]{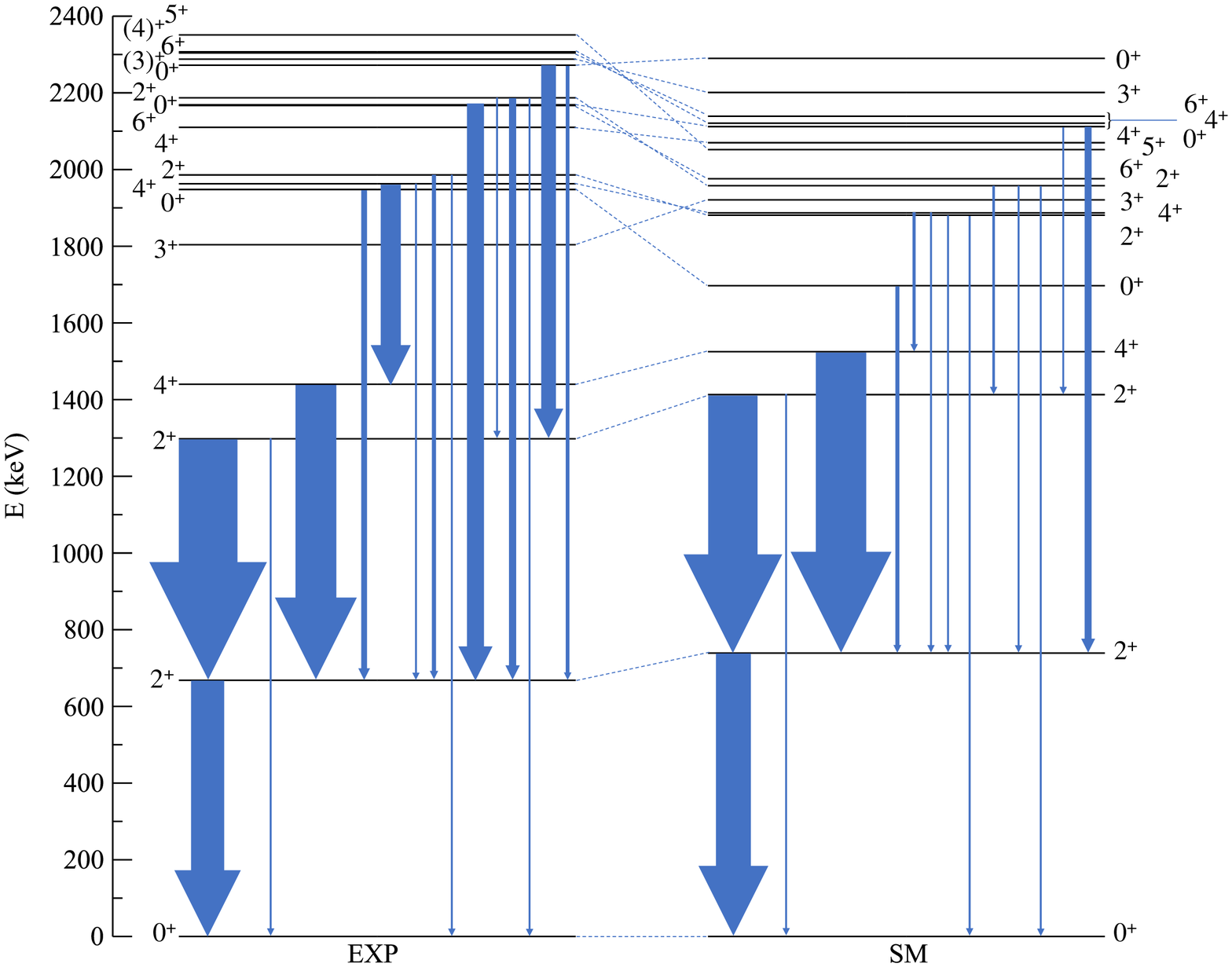}
\caption{Comparison of the experimental results with those of the present shell-model calculations using the jj55 interaction for $^{132}$Xe.  The widths of the arrows are proportional to the $B(E2)$ values.  See Table~\ref{tab:SM_E2_132Xe} for the experimental values with uncertainties.}~\label{fig:132SM}
\end{figure*}

Experimental and theoretical excitation energies and $E2$ transition rates for $^{132}$Xe are compared in Table~\ref{tab:SM_Ex_132Xe} ($E_x$s), Table~\ref{tab:SM_E2_132Xe} ($B(E2)$s) and Fig.~\ref{fig:132SM}. The dominant contributions to  the wavefunctions are listed for selected low-lying states up to $6_1^+$  in Appendix \ref{ApC}. 

Compared to $^{134}$Xe, the low-lying level sequence no longer resembles that of a vibrator. Overall, the agreement between experimental and theoretical level energies is good up to $E_x \approx 2.2$~MeV. With the exception of the 0$^+_2$, 6$^+_1$, 6$^+_2$ and 2$^+_4$ states, which are all predicted below their experimental counterparts, the calculated energies are within 100 keV of experiment. As in $^{134}$Xe, there are two predicted 6$^+$ states, close in excitation energy. In $^{132}$Xe, both states are observed and it is evident that the excitation energies of both are similarly under-predicted by the theory.  It should be noted, however, that the spectroscopic data for $^{132}$Xe is less complete than for $^{134}$Xe. For example, the yrast 8$^+$ state is yet to be identified, evidently because it occurs above the yrast 10$^+$ state, which is isomeric \cite{Vog16}. 


The overall description of the $E2$ transition rates is very good at low excitation energies, and the calculations give at least qualitative agreement with the experimental trends at higher excitation energies. The values from the calculations of Teruya \textit{et al.} \cite{Teruya.PhysRevC.92.034320} exceed the present calculations for all but the very weak $2^+_2 \rightarrow 2^+_1$ transition, a trend that probably stems from their use of $e_n = 1.0$, 20\% higher than the effective charge used here.

The spin decomposition of the wavefunctions is given along with a comparison of experimental and theoretical $g$~factors in Table~\ref{tab:spincomp132Xe}. As found for $^{134}$Xe, the present theoretical $g$~factors are in very good agreement with experiment. Those of Teruya \textit{et al.} \cite{Teruya.PhysRevC.92.034320} are also in agreement with experiment, within the experimental uncertainties. As will be discussed in more detail below, the $g$~factors can be examined along with the $E2$ strengths as a signature of the onset of collectivity, in that for collective excitations, the $g$~factors of the collective states are expected to be almost identical, with a value somewhat reduced from $g \approx Z/A \approx 0.4$. Both calculations predict $g(4^+_1) > g(2^+_1)$ consistent with experiment, $g(4^+_1) / g(2^+_1) =1.9(4)$.

\begin{table}[]
\centering
\caption{Comparison of experimental excitation energies in $^{132}$Xe with those obtained using the jj55 interaction.}
\label{tab:SM_Ex_132Xe}
\begin{ruledtabular}
\begin{tabular}{crl}
$J^{\pi}_i$  & \multicolumn{2}{c}{$E_x$ (MeV)}  \\ \cline{2-3}
  & Exp.   & SM \\
\hline
\\
0$^+_1$  & 0 & 0       \\
2$^+_1$  &  0.668 & 0.739   \\
2$^+_2$  &  1.298 & 1.413   \\
4$^+_1$  &  1.440 & 1.525    \\
3$^+_1$  &  1.804 & 1.921   \\
0$^+_2$  &  1.948 & 1.697   \\
4$^+_2$  &  1.963 & 1.887   \\
2$^+_3$  &  1.986 & 1.881   \\
4$^+_3$  &  2.110 & 2.070   \\
6$^+_1$  &  2.167 & 1.976   \\
0$^+_3$  &  2.169 & 2.112   \\
2$^+_4$  &  2.187 & 1.958   \\
0$^+_4$  &  2.272 & 2.290   \\
3$^+_2$  &  2.288 & 2.201   \\
6$^+_2$  &  2.304 & 2.139   \\
4$^+_4$  &  (2.307) & 2.121  \\
5$^+_1$  &  2.351 & 2.052   \\

\\
\end{tabular}
\end{ruledtabular}
\end{table}

\begin{table}[]
\centering
\caption{Comparison of experimental transition rates in $^{132}$Xe with those obtained using the jj55 interaction.}
\label{tab:SM_E2_132Xe}
\begin{ruledtabular}
\begin{tabular}{ccccc}
Transition  & $B(E2)_{\rm exp}$  & \multicolumn{3}{c}{$B(E2)$ ($e^2$fm$^4$)}  \\ \cline{3-5}
 & (W.u.) & Exp.\footnotemark[1]  & SM Ref.~\cite{Teruya.PhysRevC.92.034320} & Present SM\\
\hline
\\
  $2^+_1 \rightarrow 0^+_1$    & 23.1(15)           &   922(60)           &  1106    & 973  \\

  $2^+_2 \rightarrow 2^+_1$    & 41(4)              &   1640(160)         &  1490    & 1372  \\

  $2^+_2 \rightarrow 0^+_1$    & 0.079(11)          &   3.2(4)    &  0.006       &  1.1  \\

  $4^+_1 \rightarrow 2^+_1$    & 28.6(23)           &   1140(90)           &  1613   & 1401  \\

  $0^+_2 \rightarrow 2^+_1$    & 4.0$^{+31}_{-29}$  &   160(120)           &         & 105  \\

  $4^+_2 \rightarrow 4^+_1$    & 14$^{+12}_{-8}$    &  560$^{+480}_{-320}$ &         & 85.7  \\
  $4^+_2 \rightarrow 2^+_1$    & 0.45$^{+29}_{-25}$ &   18$^{+12}_{-10}$   &         & 0.075 \\

  $2^+_3 \rightarrow 2^+_2$    & 2.85$^{+92}_{-82}$  & 114$^{+37}_{-33}$  &         &  10.6  \\
  $2^+_3 \rightarrow 0^+_1$    & 1.06$^{+15}_{-13}$  & 42.3$^{+6}_{-5}$   &         &  51.2  \\

  $6^+_1 \rightarrow 4^+_1$    & 140$^{+150}_{-130}$ & 5600$^{+5800}_{-5000}$    &   1218   & 193  \\

  $0^+_3 \rightarrow 2^+_1$    & 11.7$^{+18}_{-16}$  & 467$^{+70}_{-65}$  &         & 6.8  \\

  $2^+_4 \rightarrow 2^+_2$    & 0.29$^{+66}_{-27}$   & 12$^{+26}_{-11}$   &         &  69.9  \\

  $2^+_4 \rightarrow 2^+_1$    & 5.0$^{+9}_{-8}$   & 200$^{+37}_{-33}$   &         &  38.0  \\
  \multicolumn{1}{r}{or}    & 0.27$^{+23}_{-14}$  &   11$^{+9}_{-6}$  &         &   \\
  $2^+_4 \rightarrow 0^+_1$    & 0.179$^{+28}_{25}$ &    7(1)  &         & 1.4  \\

  $0^+_4 \rightarrow 2^+_2$    & 10.2$^{+46}_{-38}$  &  407$^{+184}_{-152}$    &         & 46.5  \\
  $0^+_4 \rightarrow 2^+_1$    & 2.6$^{+10}_{-9}$  &   104$^{+40}_{-36}$  &         & 192  \\

\\
\end{tabular}
\end{ruledtabular}
\footnotetext[1]{From Ref.~\cite{NDS132} or present work.}

\end{table}

\begin{table*}[]
\centering
\caption{Spin decomposition and $g$~factors in $^{132}$Xe.}
\label{tab:spincomp132Xe}
\begin{ruledtabular}
\begin{tabular}{cllll}
$J^{\pi}_i $  & \multicolumn{3}{c}{$g$}  & Spin composition \\ \cline{2-4}
            & Exp. \cite{Xepaper.PhysRevC.65.024316} & SM Ref.~\cite{Teruya.PhysRevC.92.034320} & Present SM\\
\hline
\\
2$^+_1$  & 0.314(12) & 0.311  & 0.336  &  $ 0.43 \pi(0^+) \nu(2^+) + 0.31 \pi(2^+) \nu(0^+) + ...$ \\
2$^+_2$  & 0.1(2)    & 0.282  & 0.199  &  $ 0.39 \pi(0^+) \nu(2^+) + 0.29 \pi(2^+) \nu(2^+) + 0.10 \pi(4^+) \nu(3^+) + ...$  \\
4$^+_1$  & 0.61(11)  & 0.463  & 0.407  &  $ 0.33 \pi(2^+) \nu(2^+) + 0.23 \pi(0^+) \nu(4^+) + 0.21 \pi(4^+) \nu(0^+) + ...$   \\
4$^+_2$  &           &        & 0.648  &  $ 0.46 \pi(4^+) \nu(0^+) + 0.16 \pi(0^+) \nu(4^+) + ...$   \\
\end{tabular}
\end{ruledtabular}
\end{table*}

%
%
%

\section{Discussion}

The experimental and theoretical evidence for the emergence of collectivity in the xenon isotopes as the number of neutron holes increases from $^{136}$Xe to $^{132}$Xe is the focus of the following discussion. The shell-model calculations reported in the previous section give an overall good description of these nuclei, and at the same time, no standard collective model can account for their level schemes and electromagnetic observables.  Nevertheless, collectivity must be at least beginning to emerge in these nuclei. Here we characterize and assess the emergence of nuclear collectivity from experimental and theoretical perspectives. We will discuss the isotopes separately and then draw together an overall picture of emerging collectivity in $^{134}$Xe and $^{132}$Xe.

\subsection{$^{134}$Xe}

The vibrational-like level sequence in $^{134}$Xe has been noted and discussed previously (cf. Ref.~\cite{Pet17B} and references therein). The present calculations and comparisons between experimental and theoretical energies and electromagnetic properties show that the vibrational-like level sequence is circumstantial. The level sequence is given by a shell-model calculation with four valence protons and two valence neutron holes. The states are not fully collective admixtures of proton and neutron excitations. The experimental $g$~factor confirms the dominant proton-excitation nature of the 4$^+_1$ state, and the $E2$ transition rates from the 0$^+_2$, 2$^+_2$ and 4$^+_1$ states to the 2$^+_1$ state are not twice that of the $2^+_1 \rightarrow 0^+_1$ transition as expected for vibrational states. 

It is instructive to compare the wavefunctions of states in $^{134}$Xe (Appendix \ref{ApB}) with those of the proton states in $^{136}$Xe (Appendix \ref{ApA}) and the neutron-hole states in $^{130}$Sn, Table~\ref{tab:SM130Sn}. Looking first at the protons, it is evident that the
$\pi g_{7/2}^4$ and  $\pi g_{7/2}^2 d_{5/2}^2$ configurations remain dominant in the low-lying states of $^{134}$Xe. Turning to neutron holes, $\nu h_{11/2}^{-2}$ is dominant in $^{130}$Sn; it is still prominent in the isotone $^{134}$Xe but the $\nu d_{3/2}^{-2}$ contribution is generally stronger. With the addition of 4 protons, there is greater mixing and a redistribution of strength among the neutron partitions, which is not surprising given that the $\nu s_{1/2}$, $\nu d_{3/2}$,  and $\nu h_{11/2}$ orbitals are so close in energy.

The wavefunctions in $^{134}$Xe are becoming fragmented (cf., $^{136}$Xe and $^{128}$Sn). It is natural, therefore, to ask whether the missing $E2$ strength in the shell-model calculations  (Table~\ref{tab:SM_E2_134Xe}) is an indication of the onset of collectivity in that the size of the basis space is inadequate, or whether it is because the balance of configuration mixing in the wavefunctions is not correct. The fact that Teruya \textit{et al.} \cite{Teruya.PhysRevC.92.034320} have {\em overestimated} the $B(E2)$ strengths suggests that the difference is due to the wavefunctions, stemming from the choice of interaction, and is not an indication of the onset of collectivity (or a limitation of the basis space). The mid-shell cancellation of $E2$ strengths between seniority $\upsilon=2$ members of the $\pi g_{7/2}^4$ configuration, mentioned in relation to $^{136}$Xe above, applies in all of the Xe isotopes.  As this proton configuration remains dominant in $^{134}$Xe, the $E2$ transition strength can be strongly affected by smaller components in the wavefunction.

\subsection{$^{132}$Xe}

The comparison of wavefunctions of the states in $^{134}$Xe (Appendix \ref{ApB}) with those of the proton states in $^{136}$Xe (Appendix \ref{ApA}), and the two-neutron hole states in $^{130}$Sn (Table~\ref{tab:SM130Sn}) can be extended to $^{132}$Xe (Appendix \ref{ApC}) in this case considering the four-neutron hole states in $^{128}$Sn (Table~\ref{tab:SM128Sn}).  Looking first at the protons, it is evident that the $\pi g_{7/2}^4$ and  $\pi g_{7/2}^2 d_{5/2}^2$ configurations remain dominant in the low-excitation states of $^{132}$Xe, similar to $^{134}$Xe, but the amplitudes of the strongest components are reduced as the wavefunction becomes distributed over a much larger number of configurations.  Turning to neutron holes, the $\nu h_{11/2}^{-4}$ and $\nu h_{11/2}^{-2} d_{3/2}^{-2}$ configurations are dominant in $^{128}$Sn, but in the isotone $^{132}$Xe, the  $\nu h_{11/2}^{-2} d_{3/2}^{-2}$ configuration alone is dominant, continuing the trend that the $\nu d_{3/2}^{-2}$ configuration becomes dominant over the $\nu h_{11/2}^{-2}$ configuration in $^{134}$Xe. Nevertheless, the comparisons of the most prominent configurations in $^{132}$Xe with those in the related semimagic nuclei of $^{136}$Xe for protons and $^{128}$Sn for neutron holes, justifies the approach of Teruya \textit{et al.} \cite{Teruya.PhysRevC.92.034320}, whereby the basis states were selected by first diagonalizing over the separate proton and neutron spaces.


\subsection{Characterizing and assessing the onset of collectivity}

Is the emergence of collectivity evident in $^{134}$Xe and $^{132}$Xe?



As described above, the wavefunctions are becoming increasingly fragmented as the number of neutron holes increases. Certainly, fragmentation of the wavefunction is a requirement for the development of collective excitations; but it is not sufficient in that coherent quadrupole correlations must also be developing as the wavefunction is spread over many components. Such coherent quadrupole correlations can be measured by examining $E2$ transition strengths.

\begin{figure}[ht!]
\centering
\includegraphics[width = 7 cm, keepaspectratio, trim={0 0 0 0cm}, clip]{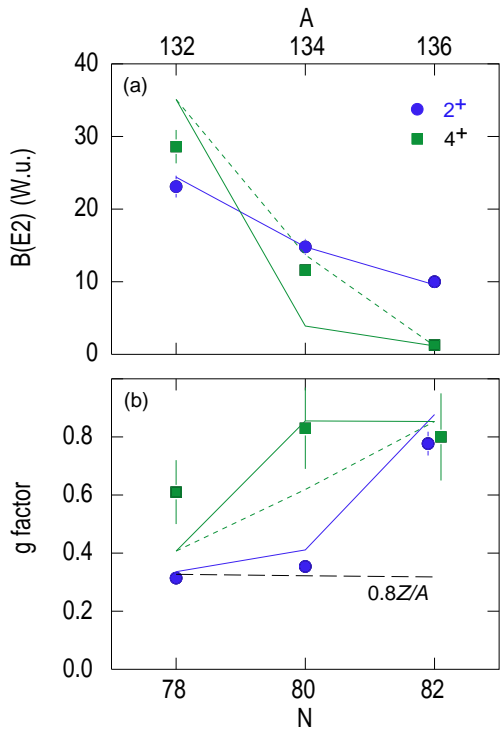}
\caption{Experimental and theoretical electromagnetic properties of $^{132}$Xe, $^{134}$Xe, and $^{136}$Xe. (a) $B(E2; 2^+_1 \rightarrow 0^+_1)$ and $B(E2; 4^+_1 \rightarrow 2^+_1)$ values. (b) $g$~factors. The data are taken from Refs. \cite{NDS132, NDS134, Pet17B, Xe136.PhysRevC.98.034302, NDS136} and present work.  In $^{134}$Xe the $E2$ transition from the 4$^+_1$ state is very sensitive to configuration mixing and some redistribution of configuration mixing with the 4$^+_2$ state in the theory is suggested. To indicate the possible impact of remixing the 4$^+$states, the solid lines indicate shell model results for the 4$^+_1$ state and the dotted lines for the 4$^+_2$ state.}~\label{fig:Xe-g-BE2}
\end{figure}

The upper panel of Fig.~\ref{fig:Xe-g-BE2} shows the experimental and theoretical $B(E2; 2^+_1 \rightarrow 0^+_1)$ and $B(E2; 4^+_1 \rightarrow 2^+_1)$ values. The $B(E2; 2^+_1 \rightarrow 0^+_1) = 10$ W.u. in $^{136}$Xe serves as a benchmark for proton single-particle strength. In the four-neutron-hole case of $^{128}$Sn, the $2^+_1 \rightarrow 0^+_1$ reduced transition strength is 4 W.u. (in experiment and theory), while for the two-neutron-hole case of $^{130}$Sn, it is 2 W.u. These values benchmark the neutron single-particle strength. It is evident that the experimental $E2$ strength in $^{134}$Xe, $15\pm 1$ W.u., already exceeds the sum of the proton and neutron parent strengths (12 W.u.). In $^{132}$Xe, the $E2$ strength of $23\pm1$ W.u., nearly doubles the sum of the proton and neutron parent strengths (14 W.u.). These trends in both theory and experiment can be interpreted as clear indicators of the emergence of collective features in the wavefunctions.

Turning to the trends in the  $4^+_1 \rightarrow 2^+_1$ transitions, the dilution of the $E2$ mid-shell cancellation associated with the seniority structure of the prominent proton configuration, $\pi g_{7/2}^4$ in $^{136}$Xe, by mixing with many other configurations, is apparent in $^{134}$Xe where the experimental $B(E2)$ is an order of magnitude higher. Theory struggles to describe the experimental $B(E2; 4^+_1 \rightarrow 2^+_1)$ value in $^{134}$Xe because of the mid-shell $E2$ cancellation of the the dominant proton configuration. Some redistribution of the configuration mixing in the theoretical 4$^+_1$ and 4$^+_2$ states is needed to explain the experimental $B(E2)$. For this reason we have indicated theoretical $B(E2)$ and $g$-factor values for both the 4$^+_1$ and 4$^+_2$ states in Fig.~\ref{fig:Xe-g-BE2}. In any case, the $E2$ strength is increasing markedly, which can be taken as an indicator of increasing collectivity.

Moving to $^{132}$Xe, in both theory and experiment,  $B(E2; 4^+_1 \rightarrow 2^+_1)$ exceeds $B(E2; 2^+_1 \rightarrow 0^+_1)$. This trend in the $4^+_1 \rightarrow 2^+_1$ transitions is also a clear signal of developing quadrupole collectivity.


The lower panel of Fig.~\ref{fig:Xe-g-BE2} shows the experimental and theoretical $g$~factors. As collectivity develops, the $g$~factors of all of the low-lying states must approach the same value near $Z/A =0.40$. Typically in collective nuclei $g \sim 0.8 Z/A$; thus for these Xe isotopes, we expect a collective $g$~factor of $g \sim 0.33$. Pronounced differences are both predicted by the shell model and observed for $g(4^+_1)/g(2_1^+)$ and $g(2^+_2)/g(2_1^+)$ in both isotopes. The 4$^+$ states retain a prominent proton contribution whereas the 2$^+_1$ states do approach $0.8 Z/A$. Thus the $E2$ transitions signal the emergence of collectivity, but the $g$~factors show that the single-particle (shell-model) structure persists in the $4^+$ and higher states.

There is no need to invoke collectivity beyond the shell-model calculations. Nevertheless, we have indicators of developing collectivity in the fragmentation of the wavefunction and the increasing $E2$ transition strengths. The picture that emerges from the shell-model calculations, looking at the fragmentation of the wavefunctions, the $E2$ transition strengths, and the $g$~factor values, is that collectivity builds up beginning with the first 2$^+$ state and then develops to higher excitation energies and spins as the number of neutron holes increases.

\section{Conclusions}

Inelastic neutron scattering was used at the University of Kentucky Accelerator Laboratory to study the nuclear structure of $^{132}$Xe.  A comprehensive level scheme was obtained, as well as new level lifetimes, multipole mixing ratios, branching ratios, and transition probabilities.  New shell-model calculations for $^{132,134,136}$Xe using \textsc{NuShellX} were also completed. The shell-model calculations account well for the level scheme and electromagnetic observables for all three isotopes. The emergence of collectivity away from the $N=82$ closed shell was evaluated by examining changes in the wavefunctions, $E2$ transition strengths, and $g$~factors as the number of neutron holes increases. The increasing complexity of the wavefunctions and the increasing $E2$ transition strengths signal emergent collectivity, whereas the $g$~factors clearly show the persistence of single-particle features in the wavefunctions for the states above the first 2$^+$ state. The picture that emerges is that collectivity builds up beginning with the first 2$^+$ state and then develops to higher excitation energies and spins as the number of neutron holes increases.

These trends are expected to develop further as more neutrons are removed, with the level structures and electromagnetic properties moving toward the patterns associated with fully collective models. At present, it is not clear whether the mass-dependent development will be slow and smooth or sudden. It will, therefore, be very useful to examine $^{130}$Xe, for which we have INS data that are currently under analysis.

\begin{acknowledgments}
This material is based upon work supported by the U.~S. National Science Foundation under Grant No. PHY-1606890.  This research was also sponsored in part by the Australian Research Council under grant No. DP170101673.  We wish to thank H. E. Baber for his valuable contributions to these measurements.
\end{acknowledgments}

\appendix
\section{Wavefunctions for $^{136}$Xe}
\label{ApA}

Table \ref{tab:SMwfn136Xe} includes additional shell-model wavefunctions of interest for $^{136}$Xe.

 \begin{table*}
 \centering
 \caption{Shell-model results for $^{136}$Xe comparing the jj55 and N82K interactions.}
 \label{tab:SMwfn136Xe}
 \begin{ruledtabular}
 \begin{tabular}{crrll}
 $J^{\pi}_i$  & \multicolumn{2}{c}{$E_x$ (MeV)}  &  Interaction & Wavefunction \\ \cline{2-3}
            & \multicolumn{1}{c}{Exp. \cite{NDS136,Xe136.PhysRevC.98.034302}} & \multicolumn{1}{c}{Theory} \\
 \hline
 \\
 0$^+_1$  & 0      & 0      &  jj55  &  $53.3\% (g_{7/2}^{4}) + 24.4\% (g_{7/2}^{2}d_{5/2}^{2}) + 11.3\% (g_{7/2}^{2} h_{11/2}^{2}) + ...$   \\
          &        & 0      &  N82K     &  $58.2\% (g_{7/2}^{4}) + 23.7\% (g_{7/2}^{2}d_{5/2}^{2}) + 6.1\% (g_{7/2}^{2} h_{11/2}^{2}) + ...$   \\
 \\
 2$^+_1$  & 1.313  & 1.329  &  jj55  &  $63.6\% (g_{7/2}^{4}) + 18.2\% (g_{7/2}^{2}d_{5/2}^{2}) + 6.4\% (g_{7/2}^{2} h_{11/2}^{2}) + ...$ \\
    &        & 1.300  &  N82K     &  $64.0\% (g_{7/2}^{4}) + 14.6\% (g_{7/2}^{2}d_{5/2}^{2}) + 9.8\% (g_{7/2}^{3} d_{3/2}) + ...$ \\
 \\
 4$^+_1$  & 1.694  & 1.660  &  jj55  &  $68.8\% (g_{7/2}^{4}) + 15.1\% (g_{7/2}^{2}d_{5/2}^{2}) + 5.8\% (g_{7/2}^{2} h_{11/2}^{2}) + ...$   \\
    &        & 1.683  &  N82K     &  $66.7\% (g_{7/2}^{4}) + 13.3\% (g_{7/2}^{2}d_{5/2}^{2}) + 7.8\% (g_{7/2}^{3} d_{5/2}^{2}) + ...$   \\
 \\
 6$^+_1$  & 1.892  & 1.809  &  jj55  &  $66.4\% (g_{7/2}^{4})+ 13.0\% (g_{7/2}^{2} d_{5/2}^{2}) + 9.6\% (g_{7/2}^{3}d_{5/2}) + ...$  \\
    &        & 1.838  &  N82K     &  $65.6\% (g_{7/2}^{4}) + 14.3\% (g_{7/2}^{3}d_{5/2}) + 10.6\% (g_{7/2}^{2} d_{5/2}^{2}) + ...$  \\
 \\
 6$^+_2$  & 2.262  & 2.022  &  jj55  &  $ 73.8\% (g_{7/2}^{3}d_{5/2}) + 9.1\% (g_{7/2}^{4}) + 6.7\% (g_{7/2} d_{5/2}^{3}) + ...$  \\
    &        & 2.199  &  N82K     &  $69.6\% (g_{7/2}^{3}d_{5/2}) + 13.5\% (g_{7/2}^{4}) + 7.8\% (g_{7/2} d_{5/2}^{3}) + ...$  \\
 \\
 0$^+_2$  & 2.581  & 2.137  &  jj55  &  $44.7\% (g_{7/2}^{2}d_{5/2}^{2}) + 35.9\% (g_{7/2}^{4}) + 11.3\% (d_{5/2}^{4} ) + ...$   \\
    &        & 2.518  &  N82K     &  $46.6\% (g_{7/2}^{2}d_{5/2}^{2}) + 28.4\% (g_{7/2}^{4}) + 9.8\% (d_{5/2}^{4} ) + ...$   \\
 \\
 4$^+_2$  & 2.156  & 2.139  &  jj55  &  $63.3\% (g_{7/2}^{4}) + 23.1\% (g_{7/2}^{3}d_{5/2}) + ...$   \\
    &        & 2.122  &  N82K     &  $78.3\% (g_{7/2}^{4}) +  9.2\% (g_{7/2}^{3}d_{5/2}) + ...$   \\
 \\
 2$^+_2$  & 2.290  & 2.229  &  jj55  &  $55.2\% (g_{7/2}^{3}d_{5/2}) + 15.6\% (g_{7/2}^{2} d_{5/2}^{2}) + 11.1\% (g_{7/2}^{4}) + ...$  \\
    &        & 2.246  &  N82K     &  $44.2\% (g_{7/2}^{3}d_{5/2}) + 34.5\% (g_{7/2}^{4}) + 5.2\% (g_{7/2}^{2} d_{5/2}^{2}) + ...$  \\
 \\
 2$^+_3$  & 2.415  & 2.357  &  jj55  &  $88.6\% (g_{7/2}^{4}) +  ...$  \\
   &        & 2.401  &  N82K     &  $48.4\% (g_{7/2}^{4}) + 31.8\% (g_{7/2}^{3}d_{5/2}) + 6.0\% (g_{7/2}^{2} d_{5/2}^{2}) + ...$  \\
 \\
 2$^+_4$  & 2.869  & 2.532  &  jj55  &  $50.1\% (g_{7/2}^{2} d_{5/2}^{2}) + 27.1\% (g_{7/2}^{3}d_{5/2}) + 5.5\% (g_{7/2}^{4}) + ...$  \\
    &        & 2.759  &  N82K     &  $51.8\% (g_{7/2}^{2} d_{5/2}^{2}) + 27.5\% (g_{7/2}^{3}d_{5/2}) + ...$  \\
 \\
 2$^+_5$  & 2.979  & 2.769  &  jj55  &  $51.0\% (g_{7/2}^{3}d_{5/2}) + 22.5\% (g_{7/2}^{2} d_{5/2}^{2}) + 5.1\% (g_{7/2}^{2} d_{5/2} d_{3/2}) + ...$  \\
    &        & 2.895  &  N82K     &  $52.5\% (g_{7/2}^{3}d_{5/2}) + 18.4\% (g_{7/2}^{2} d_{5/2}^{2}) + ...$  \\
 \\

 \end{tabular}
 \end{ruledtabular}
 \end{table*}

\section{Wavefunctions for $^{134}$Xe}
\label{ApB}
Table \ref{tab:SM134Xe_wfn} includes additional shell-model wavefunctions of interest for $^{134}$Xe.
\begin{table*}
\centering
\caption{Shell-model wavefunctions for $^{134}$Xe using the jj55 interaction.}
\label{tab:SM134Xe_wfn}
\begin{ruledtabular}
\begin{tabular}{cl}
$J^{\pi}_i$   & \multicolumn{1}{c}{Wavefunction} \\
\hline
\\
0$^+_1$  &  $43.6\% \left [ \pi (g_{7/2}^{4})_{0^+} \otimes \nu [ 0.43(d_{3/2}^{-2})_{0^+} + 0.25(h_{11/2}^{-2})_{0^+} + 0.17(s_{1/2}^{-2})_{0^+} + ... ] \right ] + $\\
         &  $18.9\% \left [ \pi (g_{7/2}^{2}d_{5/2}^{2})_{0^+} \otimes \nu [ 0.37(h_{11/2}^{-2})_{0^+} + 0.32(d_{3/2}^{-2})_{0^+} + 0.15(s_{1/2}^{-2})_{0^+} + ... ] \right ] + $\\
         &  $9.4\% \left [ \pi (g_{7/2}^{2}h_{11/2}^{2})_{0^+} \otimes \nu [ 0.38(d_{3/2}^{-2})_{0^+} + 0.30(h_{11/2}^{-2})_{0^+} + 0.16(s_{1/2}^{-2})_{0^+} + ... ] \right ] + $\\
         &  $8.3\% \left [ \pi (g_{7/2}^{4})_{2^+} \otimes \nu [ 0.21(d_{3/2}^{-2})_{2^+} + 0.20(s_{1/2}^{-1} d_{3/2}^{-1})_{2^+}  + 0.17(h_{11/2}^{-2})_{2^+} + ... ] \right ] + ...$\\
\\
2$^+_1$  &  $26.8\% \left [ \pi (g_{7/2}^{4})_{0^+} \otimes \nu [ 0.39(s_{1/2}^{-1} d_{3/2}^{-1})_{2^+} + 0.25 (d_{3/2}^{-2})_{2^+} +... ] \right ] +$\\
         &  $22.6\% \left [ \pi (g_{7/2}^{4})_{2^+} \otimes \nu [ 0.46(d_{3/2}^{-2})_{0^+} + 0.20(s_{1/2}^{-2})_{0^+} + 0.20(h_{11/2}^{-2})_{0^+} +... ] \right ] +$\\
         &  $9.3\% \left [ \pi (g_{7/2}^{2}d_{5/2}^{2})_{0^+} \otimes \nu [ 0.35(s_{1/2}^{-1} d_{3/2}^{-1})_{2^+} + 0.20(h_{11/2}^{-2})_{2^+} + ... ] \right ] +$\\
         &  $7.2\% \left [ \pi (g_{7/2}^{2}d_{5/2}^{2})_{2^+} \otimes \nu [ 0.34(d_{3/2}^{-2})_{0^+} + 0.31(h_{11/2}^{-2})_{0^+} +... ] \right ] + ... $\\
 \\
0$^+_2$  &  $27.0\% \left [ \pi (g_{7/2}^{4})_{0^+} \otimes \nu [ 0.55(d_{3/2}^{-2})_{0^+} + 0.24(s_{1/2}^{-2})_{0^+} + 0.19(h_{11/2}^{-2})_{0^+} +... ] \right ] + $\\
         &  $24.5\% \left [ \pi (g_{7/2}^{2}d_{5/2}^{2})_{0^+} \otimes \nu [ 0.90(h_{11/2}^{-2})_{0^+} + ... ] \right ] + $\\
         &  $15.3\% \left [ \pi (g_{7/2}^{4})_{2^+} \otimes \nu [ 0.40(s_{1/2}^{-1} d_{3/2}^{-1})_{2^+}  + 0.29(d_{3/2}^{-2})_{2^+} + 0.12 (s_{1/2}^{-1} d_{5/2}^{-1})_{2^+} + ... ] \right ] + $\\
         &  $5.2\% \left [ \pi (g_{7/2}^{2}h_{11/2}^{2})_{0^+} \otimes \nu [ 0.72(h_{11/2}^{-2})_{0^+} + ... ] \right ] + ...$\\
 \\
2$^+_2$  &  $18.9\% \left [ \pi (g_{7/2}^{4})_{0^+} \otimes \nu [ 0.55(s_{1/2}^{-1} d_{3/2}^{-1})_{2^+} + 0.24 (d_{3/2}^{-2})_{2^+} +... ] \right ] +$\\
         &  $18.1\% \left [ \pi (g_{7/2}^{4})_{2^+} \otimes \nu [ 0.41(h_{11/2}^{-2})_{0^+} + 0.32(d_{3/2}^{-2})_{0^+} + ... ] \right ] +$\\
         &  $8.2\% \left [ \pi (g_{7/2}^{2}d_{5/2}^{2})_{2^+} \otimes \nu [ 0.63(h_{11/2}^{-2})_{0^+} + 0.16(d_{3/2}^{-2})_{0^+} + ... ] \right ] +$\\
         &  $5.5\% \left [ \pi (g_{7/2}^{2}d_{5/2}^{2})_{0^+} \otimes \nu [ 0.42(s_{1/2}^{-1} d_{3/2}^{-1})_{2^+} + 0.34(h_{11/2}^{-2})_{2^+} + ... ] \right ] + ... $\\
 \\
4$^+_1$  &  $47.3\% \left [ \pi (g_{7/2}^{4})_{4^+} \otimes \nu [ 0.44(d_{3/2}^{-2})_{0^+} + 0.24(h_{11/2}^{-2})_{0^+} + 0.17(s_{1/2}^{-2})_{0^+} + ... ] \right ] +$\\
         &  $11.4\% \left [ \pi (g_{7/2}^{2}d_{5/2}^{2})_{4^+} \otimes \nu [ 0.39(h_{11/2}^{-2})_{0^+} + 0.31(d_{3/2}^{-2})_{0^+} + 0.14(s_{1/2}^{-2})_{0^+} + ... ] \right ] +$\\
         &  $6.4\% \left [ \pi (g_{7/2}^{3}d_{5/2})_{4^+} \otimes \nu [ 0.35(d_{3/2}^{-2})_{0^+} + 0.33(h_{11/2}^{-2})_{0^+} + 0.16(s_{1/2}^{-2})_{0^+} + ... ] \right ] + ... $\\
\\
4$^+_2$  &  $38.2\% \left [ \pi (g_{7/2}^{4})_{4^+} \otimes \nu [ 0.47(d_{3/2}^{-2})_{0^+} + 0.19(s_{1/2}^{-2})_{0^+} + 0.19(h_{11/2}^{-2})_{0^+} +... ] \right ] +$\\
         &  $19.2\% \left [ \pi (g_{7/2}^{4})_{4^+} \otimes \nu [ 0.39(s_{1/2}^{-1} d_{3/2}^{-1})_{2^+}  + 0.24(d_{3/2}^{-2})_{2^+} + 0.12(s_{1/2}^{-1} d_{5/2}^{-1})_{2^+} +... ] \right ] +$\\
         &  $6.8\% \left [ \pi (g_{7/2}^{4}d_{5/2})_{0^+} \otimes \nu [ 0.65(h_{11/2}^{-2})_{4^+}   + 0.14(d_{3/2}^{-1} d_{5/2}^{-1})_{4^+} + ... ] \right ] + ... $\\
\\
6$^+_1$  &  $36.4\% \left [ \pi (g_{7/2}^{4})_{6^+} \otimes \nu [ 0.44(d_{3/2}^{-2})_{0^+} + 0.24(h_{11/2}^{-2})_{0^+} + 0.17(s_{1/2}^{-2})_{0^+} + ... ] \right ] +$\\
         &  $24.4\% \left [ \pi (g_{7/2}^{3}d_{5/2})_{6^+} \otimes \nu [ 0.34(d_{3/2}^{-2})_{0^+} + 0.34(h_{11/2}^{-2})_{0^+} + 0.16(s_{1/2}^{-2})_{0^+} + ... ] \right ] +$\\
         &  $7.3\% \left [ \pi (g_{7/2}^{2}d_{5/2}^{2})_{6^+} \otimes \nu [ 0.39(h_{11/2}^{-2})_{0^+} + 0.31(d_{3/2}^{-2})_{0^+} + 0.14(s_{1/2}^{-2})_{0^+} + ... ] \right ] + ...  $\\
\\
6$^+_2$  &  $37.8\% \left [ \pi (g_{7/2}^{3}d_{5/2})_{6^+} \otimes \nu [ 0.35(d_{3/2}^{-2})_{0^+} + 0.33(h_{11/2}^{-2})_{0^+} + 0.16(s_{1/2}^{-2})_{0^+} + ... ] \right ] + $\\
         &  $24.6\% \left [ \pi (g_{7/2}^{4})_{6^+} \otimes \nu [ 0.45(d_{3/2}^{-2})_{0^+} + 0.22(h_{11/2}^{-2})_{0^+} + 0.17(s_{1/2}^{-2})_{0^+} + ... ] \right ] + ... $\\
\\

\end{tabular}
\end{ruledtabular}
\end{table*}

\section{Wavefunctions for $^{132}$Xe}
\label{ApC}
Table \ref{tab:SM132Xe_wfn} includes additional shell-model wavefunctions of interest for $^{132}$Xe.
\begin{table*}
\centering
\caption{Shell-model wavefunctions for $^{132}$Xe using the jj55 interaction.}
\label{tab:SM132Xe_wfn}
\begin{ruledtabular}
\begin{tabular}{ll}
$J^{\pi}_i$  & Wavefunction \\
\hline
\\
0$^+_1$  &  $31.4\% \left [ \pi (g_{7/2}^{4})_{0^+} \otimes \nu [ 0.35(h_{11/2}^{-2} d_{3/2}^{-2})_{0^+} + 0.11(h_{11/2}^{-2} s_{1/2}^{-2})_{0^+} + 0.09 (s_{1/2}^{-2} d_{3/2}^{-2})_{0^+} +  0.07 (h_{11/2}^{-2} d_{5/2}^{-2})_{0^+} + \right . $\\
         &  $\left . 0.07 (d_{3/2}^{-2} d_{5/2}^{-2})_{0^+} + 0.06(h_{11/2}^{-4})_{0^+} + 0.05 (h_{11/2}^{-2} g_{7/2}^{-2})_{0^+}  +
         0.05 (d_{3/2}^{-4})_{0^+} + 0.04(d_{3/2}^{-2} g_{7/2}^{-2})_{0^+} + ... ] \right ] + $\\
         &  $17.4\% \left [ \pi (g_{7/2}^{2}d_{5/2}^{2})_{0^+} \otimes \nu [ 0.35(h_{11/2}^{-2} d_{3/2}^{-2})_{0^+} + 0.13(h_{11/2}^{-2} s_{1/2}^{-2})_{0^+} + 0.11(h_{11/2}^{-4})_{0^+} +  0.08 (h_{11/2}^{-2} d_{5/2}^{-2})_{0^+} + \right . $\\
         &  $\left . 0.07 (h_{11/2}^{-2} g_{7/2}^{-2})_{0^+}  + ... ] \right ] +
         11.0\% \left [ \pi (g_{7/2}^{4})_{2^+} \otimes \nu [ 0.29(h_{11/2}^{-2} d_{3/2}^{-2})_{2^+} + 0.11(h_{11/2}^{-2} s_{1/2}^{-1}d_{3/2}^{-1})_{2^+} +  ... ] \right ] +$\\
         &  $2.8\% \left [ \pi (g_{7/2}^{2}h_{11/2}^{2})_{0^+} \otimes \nu [ (h_{11/2}^{-2} d_{3/2}^{-2})_{0^+}  + ... ] \right ] +
         1.7\% \left [ \pi (g_{7/2}^{2}d_{5/2}^{2})_{2^+} \otimes \nu [ (h_{11/2}^{-2} d_{3/2}^{-2})_{0^+} + ... ] \right ] +$\\
         &  $1.4\% \left [ \pi (g_{7/2}^{2}d_{3/2}^{2})_{0^+} \otimes \nu [ (h_{11/2}^{-2} d_{3/2}^{-2})_{0^+}  + ... ] \right ] + ... $\\
\\
2$^+_1$  &  $19.6\% \left [ \pi (g_{7/2}^{4})_{0^+} \otimes \nu [ 0.33(h_{11/2}^{-2} d_{3/2}^{-2})_{2^+} +  0.13(h_{11/2}^{-2} s_{1/2}^{-1} d_{3/2}^{-1})_{2^+}  +
                            0.06(h_{11/2}^{-2} s_{1/2}^{-2})_{2^+} + ... ] \right ] +$\\
         &  $13.1\% \left [ \pi (g_{7/2}^{4})_{2^+} \otimes \nu [ 0.37(h_{11/2}^{-2} d_{3/2}^{-2})_{0^+} + 0.12(h_{11/2}^{-2} s_{1/2}^{-2})_{0^+} +  0.09(s_{1/2}^{-2} d_{3/2}^{-0})_{2^+} +... ] \right ] +$\\
         &  $11.5\% \left [ \pi (g_{7/2}^{2}d_{5/2}^{2})_{0^+} \otimes \nu [ 0.32(h_{11/2}^{-2} d_{3/2}^{-2})_{2^+} + 0.14(h_{11/2}^{-2} s_{1/2}^{-1} d_{3/2}^{-1})_{2^+}  + ... ] \right ] +$\\
         &  $7.4\% \left [ \pi (g_{7/2}^{2}d_{5/2}^{2})_{2^+} \otimes \nu [ 0.36(h_{11/2}^{-2} d_{3/2}^{-2})_{0^+} + 0.14(h_{11/2}^{-2} s_{1/2}^{-2})_{0^+} + ... ] \right ] + ... $\\
 \\
0$^+_2$  &  $26.8\% \left [ \pi (g_{7/2}^{4})_{0^+} \otimes \nu [ 0.33(s_{1/2}^{-2} d_{3/2}^{-2})_{0^+} +  0.18(h_{11/2}^{-2} d_{3/2}^{-2})_{0^+} + \right . $\\
         &  $\left . 0.11 (d_{3/2}^{-2} d_{5/2}^{-2})_{0^+} + 0.08(h_{11/2}^{-2} s_{1/2}^{-2})_{0^+} + 0.05 (d_{3/2}^{-4})_{0^+} +
         0.04(d_{3/2}^{-2} g_{7/2}^{-2})_{0^+} + ... ] \right ] + $\\
         &  $19.3\% \left [ \pi (g_{7/2}^{2}d_{5/2}^{2})_{0^+} \otimes \nu [ 0.33(h_{11/2}^{-2} d_{3/2}^{-2})_{0^+} + 0.22(h_{11/2}^{-4})_{0^+} + 0.11(h_{11/2}^{-2} s_{1/2}^{-2})_{0^+} + 0.11 (h_{11/2}^{-2} g_{7/2}^{-2})_{0^+}  +  \right . $\\
         &  $\left . 0.10 (h_{11/2}^{-2} d_{5/2}^{-2})_{0^+} + ... ] \right ] + $\\
         &  $12.9\% \left [ \pi (g_{7/2}^{2}d_{5/2}^{2})_{2^+} \otimes \nu [ 0.27(h_{11/2}^{-2} d_{3/2}^{-2})_{2^+} + 0.16 (h_{11/2}^{-2} s_{1/2}^{-1}d_{3/2}^{-1})_{2^+} +
         0.13(h_{11/2}^{-4})_{2^+} +  ... ] \right ] +$\\
         &  $12.2\% \left [ \pi (g_{7/2}^{4})_{2^+} \otimes \nu [ 0.15(h_{11/2}^{-2} d_{3/2}^{-2})_{2^+} + 0.12(s_{1/2}^{-1} d_{3/2}^{-2} d_{5/2}^{-1})_{2^+} + 0.11(s_{1/2}^{-1} d_{3/2}^{-3})_{2^+} +
         0.09(h_{11/2}^{-2} s_{1/2}^{-1}d_{3/2}^{-1})_{2^+} +  \right . $\\
         & $\left .  0.09(s_{1/2}^{-2} d_{3/2}^{-2})_{2^+} + ... ] \right ] +
         1.2\% \left [ \pi (g_{7/2}^{2}d_{5/2} s_{1/2})_{2^+} \otimes \nu [(h_{11/2}^{-2} d_{3/2}^{-2})_{2^+} + ... ] \right ] + ... $\\
\\
2$^+_2$  &  $17.1\% \left [ \pi (g_{7/2}^{4})_{0^+} \otimes \nu [ 0.31(h_{11/2}^{-2} s_{1/2}^{-1} d_{3/2}^{-1})_{2^+}  + 0.29(h_{11/2}^{-2} d_{3/2}^{-2})_{2^+}  +
                           0.08(h_{11/2}^{-2} d_{3/2}^{-1} g_{7/2}^{-1})_{2^+} + 0.07(h_{11/2}^{-2} s_{1/2}^{-1} d_{5/2}^{-1})_{2^+} + ... ] \right ] +$\\
         &  $10.9\% \left [ \pi (g_{7/2}^{2}d_{5/2}^{2})_{0^+} \otimes \nu [ 0.32(h_{11/2}^{-2} s_{1/2}^{-1} d_{3/2}^{-1})_{2^+}  + 0.27(h_{11/2}^{-2} d_{3/2}^{-2})_{2^+}+ ... ] \right ] +$\\
         &  $10.9\% \left [ \pi (g_{7/2}^{4})_{2^+} \otimes \nu [ 0.35(h_{11/2}^{-2} d_{3/2}^{-2})_{2^+} + 0.16(h_{11/2}^{-2} s_{1/2}^{-1} d_{3/2}^{-1})_{2^+}  +  ... ] \right ] +$\\
         &  $7.0\% \left [ \pi (g_{7/2}^{2}d_{5/2}^{2})_{2^+} \otimes \nu [ 0.35(h_{11/2}^{-2} d_{3/2}^{-2})_{2^+} + 0.14(h_{11/2}^{-2} s_{1/2}^{-1} d_{3/2}^{-1})_{2^+}  + ... ] \right ] +$\\
         &  $4.5\% \left [ \pi (g_{7/2}^{2}h_{11/2}^{2})_{0^+} \otimes \nu [ 0.33(h_{11/2}^{-2} s_{1/2}^{-1} d_{3/2}^{-1})_{2^+}  + 0.28(h_{11/2}^{-2} d_{3/2}^{-2})_{2^+}+ ... ] \right ] +$\\
         &  $1.5\% \left [ \pi (g_{7/2}^{4})_{2^+} \otimes \nu [(h_{11/2}^{-2} s_{1/2}^{-1} d_{3/2}^{-1})_{3^+} +... ] \right ] + ... $\\
 \\
4$^+_1$  &  $12.5\% \left [ \pi (g_{7/2}^{4})_{2^+} \otimes \nu [ 0.38(h_{11/2}^{-2} d_{3/2}^{-2})_{2^+} + 0.11(h_{11/2}^{-2} s_{1/2}^{-1} d_{3/2}^{-1})_{2^+} +  ... ] \right ] +$\\
         &  $9.9\% \left [ \pi (g_{7/2}^{4})_{0^+} \otimes \nu [ 0.37(h_{11/2}^{-2} d_{3/2}^{-2})_{4^+} + 0.15(h_{11/2}^{-2} s_{1/2}^{-2})_{4^+} +  ... ] \right ] +$\\
         &  $3.0\% \left [ \pi (g_{7/2}^{4})_{4^+} \otimes \nu [ (h_{11/2}^{-2} d_{3/2}^{-2})_{0^+} + ... ] \right ] +
         2.7\% \left [ \pi (g_{7/2}^{2}d_{5/2}^{2})_{2^+} \otimes \nu [ (h_{11/2}^{-2} d_{3/2}^{-2})_{2^+} +... ] \right ] +$\\
         &  $2.2\% \left [ \pi (g_{7/2}^{2}d_{5/2}^{2})_{0^+} \otimes \nu [ (h_{11/2}^{-2} d_{3/2}^{-2})_{4^+}+ ... ] \right ] +
         1.4\% \left [ \pi (g_{7/2}^{2}d_{5/2}^{2})_{4^+} \otimes \nu [ (h_{11/2}^{-2} d_{3/2}^{-2})_{0^+}+ ... ] \right ] + $\\
         &  $1.2\% \left [ \pi (g_{7/2}^{3}d_{5/2})_{4^+} \otimes \nu [ (h_{11/2}^{-2} d_{3/2}^{-2})_{0^+} +... ] \right ]  +
         1.0\% \left [ \pi (g_{7/2}^{2}d_{3/2}^{2})_{2^+} \otimes \nu [ (h_{11/2}^{-2} d_{3/2}^{-2})_{2^+} +... ] \right ]  +... $\\
 \\
4$^+_2$  &  $21.8\% \left [ \pi (g_{7/2}^{4})_{4^+} \otimes \nu [ 0.35(h_{11/2}^{-2} d_{3/2}^{-2})_{0^+} + 0.11(h_{11/2}^{-2} s_{1/2}^{-2})_{0^+} + 0.10(s_{1/2}^{-2} d_{3/2}^{-2})_{0^+} + 0.07(d_{3/2}^{-2} d_{5/2}^{-2})_{0^+} +  \right . $\\
         &  $\left . 0.06(h_{11/2}^{-4})_{0^+}  + 0.06(h_{11/2}^{-2} d_{5/2}^{-2})_{0^+} + 0.05(h_{11/2}^{-2} g_{7/2}^{-2})_{0^+} + ... ] \right ] +
         3.5\% \left [ \pi (g_{7/2}^{4})_{0^+} \otimes \nu [ (h_{11/2}^{-2} d_{3/2}^{-2})_{4^+} + ... ] \right ] + $\\
         &  $2.7\% \left [ \pi (g_{7/2}^{2}d_{5/2}^{2})_{4^+} \otimes \nu [ (h_{11/2}^{-2} d_{3/2}^{-2})_{0^+} +... ] \right ] +
         2.0\% \left [ \pi (g_{7/2}^{3}d_{5/2})_{4^+} \otimes \nu [ (h_{11/2}^{-2} d_{3/2}^{-2})_{0^+} +... ] \right ] +$\\
         & $1.7\% \left [ \pi (g_{7/2}^{2}d_{5/2}^{2})_{0^+} \otimes \nu [ (h_{11/2}^{-2} d_{3/2}^{-2})_{4^+}+ ... ] \right ] +
         1.2\% \left [ \pi (g_{7/2}^{4})_{4^+} \otimes \nu [ (h_{11/2}^{-2} d_{3/2}^{-2})_{2^+}+ ... ] \right ] + ... $\\
 \\
 6$^+_1$ &  $40.1\% \left [ \pi (g_{7/2}^{3}d_{5/2})_{6^+} \otimes \nu [
          0.36(h_{11/2}^{-2} d_{3/2}^{-2})_{0^+} + 0.13 (h_{11/2}^{-2} s_{1/2}^{-2})_{0^+} + 0.08 (h_{11/2}^{-4})_{0^+} +
          0.07 (h_{11/2}^{-2} d_{5/2}^{-2})_{0^+} + 0.07(d_{3/2}^{-2} d_{5/2}^{-2})_{0^+} +  \right . $ \\
          & $ \left . 0.06 (h_{11/2}^{-2} g_{7/2}^{-2})_{0^+} + 0.06 (s_{1/2}^{-2} d_{3/2}^{-2})_{0^+} +
          0.03 (d_{3/2}^{-2} g_{7/2}^{-2})_{0^+} +          ... ] \right ] + $\\
         &  $6.5\% \left [ \pi (g_{7/2}^{3}d_{5/2})_{5^+} \otimes \nu [ 0.26 (h_{11/2}^{-2} d_{3/2}^{-2})_{2^+} + 0.18 (h_{11/2}^{-2} s_{1/2}^{-1} d_{3/2}^{-1})_{2^+}  + ...] \right ] + $ \\
         &  $1.6\% \left [ \pi (g_{7/2}^{3}d_{5/2})_{4^+} \otimes \nu [ (h_{11/2}^{-2} d_{3/2}^{-2})_{2^+} + ...] \right ] +
         1.3\% \left [ \pi (g_{7/2}^{3}d_{5/2})_{8^+}\otimes \nu [ (h_{11/2}^{-2} d_{3/2}^{-2})_{0^+} +... ] \right ] + $\\
         &  $1.0\% \left [ \pi (g_{7/2}d_{5/2}^{3})_{6^+} \otimes \nu [ (h_{11/2}^{-2} d_{3/2}^{-2})_{0^+} +... ] \right ] + ... $\\
 \\

\end{tabular}
\end{ruledtabular}
\end{table*}

\clearpage

\bibliography{132Xe}

 \end{document}